%% file: main.tex
\documentclass[sigconf,nonacm,screen]{acmart}

\usepackage{microtype}
\usepackage{graphicx}
\usepackage{booktabs} %

\usepackage{hyperref}
\usepackage{makecell}
\usepackage{multirow}
\usepackage{adjustbox} 
\usepackage[table]{xcolor}

\definecolor{brandblue}{RGB}{33, 150, 243}
\colorlet{accent}{brandblue}

\usepackage{outlines}
\usepackage{listings}
\usepackage{float}
\usepackage{tikz}
\usepackage{enumitem}
\usepackage{caption}
\usepackage{subcaption}

\setlist[enumerate]{nosep,leftmargin=*}
\setlist[itemize]{nosep,leftmargin=*}

\definecolor{codebg}{RGB}{248,248,248}
\definecolor{codecomment}{RGB}{128,128,128}
\definecolor{codekeyword}{RGB}{0,92,197}
\definecolor{codestring}{RGB}{163,21,21}
\definecolor{codenumber}{RGB}{120,120,120}

\definecolor{codegreen}{rgb}{0,0.6,0}
\definecolor{codegray}{rgb}{0.5,0.5,0.5}
\definecolor{codepurple}{rgb}{0.58,0,0.82}
\definecolor{backcolour}{rgb}{0.95,0.95,0.92}
\definecolor{textblue}{rgb}{.2,.2,.7}
\definecolor{textred}{rgb}{0.54,0,0}
\definecolor{textgreen}{rgb}{0,0.43,0}
\definecolor{codered}{rgb}{201,72,12}

\lstdefinestyle{mintedlike}{
    basicstyle=\linespread{0.9}\ttfamily\footnotesize,
    commentstyle=\color{codecomment},
    showstringspaces=false,
    showspaces=false,
    captionpos=b,
}

\lstdefinestyle{syncopate}{
language=Python,
basicstyle=\linespread{0.9}\ttfamily,
numbers=left,
numberstyle=\tiny, 
stepnumber=1,
numbersep=5pt, 
keywordstyle=\bfseries\color{codegreen},
commentstyle=\color{textred},   
stringstyle=\color{textgreen},
columns=fullflexible,
keepspaces=true,
xleftmargin=\parindent,
showstringspaces=false,
otherkeywords = {True, False},
keywordstyle=[2]\color{codepurple}\bfseries,
keywordstyle=[3]\color{textblue}\bfseries,
keywords=[2]{Tensor, Callable},
keywords=[3]{__init__, forward},
keywordstyle=[4]\color{codegreen},
keywords=[4]{self},
captionpos=b,
}
\input{macro}

\begin{document}

\title{\sysname: A Kernel Overlapping Framework with PyTorch-like Modularity and Megakernel Performance} %

\author{Jinkun Lin}
    \affiliation{\institution{Cornell University}\city{Ithaca}\state{NY}\country{USA}}
\author{Byungsoo Oh}
    \affiliation{\institution{Cornell University}\city{Ithaca}\state{NY}\country{USA}}
\author{Rachee Singh}
    \affiliation{\institution{Cornell University}\city{Ithaca}\state{NY}\country{USA}}
\begin{abstract}

Modern ML frameworks execute models using the \kernelperop paradigm, which runs each operator as a separate GPU kernel. This paradigm is modular, since developers can compose various kernel implementations and profile each kernel in isolation. However, this paradigm enforces dependencies at the granularity of entire kernels, which places a barrier between dependent kernels. These barriers cause streaming multiprocessors to idle, serialize computation and communication, and prevent a kernel from prefetching its weights while the preceding kernel runs. Megakernels remove kernel barriers by fusing the operators of an entire layer or model into a single kernel, but they give up the modularity of the \kernelperop paradigm. In this work, we ask whether a \kernelperop ML framework can perform on par with megakernels. Our key insight is that removing kernel barriers does not require fusing kernels. We develop \sysname, a framework that launches dependent kernels on separate CUDA streams and enforces each dependency on individual tiles of the intermediate tensor that the producer kernel writes and the consumer kernel reads. Since kernels synchronize on data tiles instead of on each other's thread blocks, any two kernels that adopt the same data tiling can be composed without modifying either kernel. We evaluate \sysname on dense Llama3 and Qwen3 models on single H100, A100, and L40 GPUs and on multi-GPU servers with $4\times$A100 and $8\times$H100. Our \sysname framework matches the throughput of hand-tuned ThunderKittens megakernels, and achieves 1.09$\times$--1.10$\times$ the decode throughput and 0.98$\times$--1.10$\times$ the prefill throughput of the best-performing baseline, while reusing existing Triton, cuBLAS, and cuDNN kernels.

\end{abstract}

\date{}
\maketitle

\input{sections/intro-v2}

\input{sections/method}
\input{sections/impl}

\input{sections/eval}
\input{sections/related-work}

\input{sections/conclusion}

\bibliographystyle{plain}
\bibliography{bib}

\input{sections/appendix}

\end{document}

%% file: macro.tex
\usepackage{xspace}
\usepackage{xcolor}
\usepackage[normalem]{ulem}

\newcommand{\allnotes}[1]{}
\newcommand{\ignore}[1]{}

\newcommand{\jk}[1]{\allnotes{\textcolor{orange}{[Jinkun: #1]}}}

\newcommand{\eg}{\textit{e.g.,}\xspace}
\newcommand{\ie}{\textit{i.e.,}\xspace}
\providecommand{\vs}{vs.\@\xspace}

\newcommand{\sysname}{MegaOps\xspace}

\newcommand{\kernelperop}{kernel-per-operator\xspace}

\newcommand{\TK}{ThunderKittens\xspace}

\newcommand{\ETC}{Event Tensor\xspace}

\input{numbers/speedups_1xH100-short}

\input{numbers/speedups_1xH100-long}

\input{numbers/speedups_8xH100-long}

\input{numbers/speedups_1xA100-SXM4-40GB-short}

\input{numbers/speedups_1xA100-SXM4-40GB-long}

\input{numbers/speedups_4xA100-SXM4-40GB-long}

\input{numbers/speedups_4xA100-SXM4-40GB-short}
\input{numbers/speedups_1xL40-short}

\input{numbers/mbu_a100_short-decode-1gpu}
\input{numbers/mbu_l40_short-decode-1gpu}
\providecommand{\myparab}[1]{\vspace{0pt}\noindent\textbf{#1} }

\newenvironment{packeditemize}{\begin{list}{$\bullet$}{\setlength{\itemsep}{0.1pt}\addtolength{\labelwidth}{0pt}\setlength{\leftmargin}{\labelwidth}\setlength{\listparindent}{\parindent}\setlength{\parsep}{1pt}\setlength{\topsep}{0pt}}}{\end{list}}

%% file: numbers/speedups_1xH100-short.tex
\newcommand{\hSingleShortDecodeOursVsVllmMin}{1.14}
\newcommand{\hSingleShortDecodeOursVsVllmMax}{1.48}
\newcommand{\hSingleShortDecodeOursVsVllmAvg}{1.27}

\newcommand{\hSingleShortDecodeOursVsNoovlpAvg}{1.21}

\newcommand{\hSingleShortDecodeNoovlpVsVllmAvg}{1.05}

\newcommand{\hSingleShortDecodeOursVsVDCoresMin}{0.96}
\newcommand{\hSingleShortDecodeOursVsVDCoresMax}{1.09}

\newcommand{\hSingleShortDecodeOursVsTKAvg}{0.89}

%% file: numbers/speedups_1xH100-long.tex
\newcommand{\hSingleLongDecodeOursVsVllmMin}{1.07}
\newcommand{\hSingleLongDecodeOursVsVllmMax}{1.40}
\newcommand{\hSingleLongDecodeOursVsVllmAvg}{1.19}

\newcommand{\hSingleLongDecodeOursVsTKAvg}{1.33}

%% file: numbers/speedups_8xH100-long.tex
\newcommand{\hMultiLongPrefillAGRSFromBSQwenEight}{8}
\newcommand{\hMultiLongPrefillAGRSFromBSLlamaSeventy}{2}

\newcommand{\hMultiLongDecodeOursVsVllmMin}{1.01}
\newcommand{\hMultiLongDecodeOursVsVllmMax}{1.20}
\newcommand{\hMultiLongDecodeOursVsVllmAvg}{1.11}

\newcommand{\hMultiLongPrefillOursVsVllmMin}{1.01}
\newcommand{\hMultiLongPrefillOursVsVllmMax}{1.48}
\newcommand{\hMultiLongPrefillOursVsVllmAvg}{1.18}

\newcommand{\hMultiLongPrefillOursVsNoovlpAvg}{1.24}

\newcommand{\hMultiLongPrefillOursVsTKMin}{0.94}
\newcommand{\hMultiLongPrefillOursVsTKMax}{1.15}

\newcommand{\hMultiLongPrefillOursAGRSVsOursARBSOneMin}{0.72}
\newcommand{\hMultiLongPrefillOursAGRSVsOursARBSOneMax}{0.90}

\newcommand{\hMultiLongPrefillOursAGRSVsOursARBSSixteenMin}{1.08}
\newcommand{\hMultiLongPrefillOursAGRSVsOursARBSSixteenMax}{1.13}

%% file: numbers/speedups_1xA100-SXM4-40GB-short.tex
\newcommand{\aSingleShortDecodeOursVsVllmMin}{1.10}
\newcommand{\aSingleShortDecodeOursVsVllmMax}{1.48}
\newcommand{\aSingleShortDecodeOursVsVllmAvg}{1.23}

\newcommand{\aSingleShortDecodeOursVsNoovlpAvg}{1.10}

\newcommand{\aSingleShortDecodeNoovlpVsVllmAvg}{1.12}

%% file: numbers/speedups_1xA100-SXM4-40GB-long.tex
\newcommand{\aSingleLongDecodeOursVsVllmMin}{1.03}
\newcommand{\aSingleLongDecodeOursVsVllmMax}{1.38}
\newcommand{\aSingleLongDecodeOursVsVllmAvg}{1.17}

%% file: numbers/speedups_4xA100-SXM4-40GB-long.tex
\newcommand{\aMultiLongDecodeOursVsVllmMin}{1.10}
\newcommand{\aMultiLongDecodeOursVsVllmMax}{1.20}
\newcommand{\aMultiLongDecodeOursVsVllmAvg}{1.15}

\newcommand{\aMultiLongPrefillOursVsVllmMin}{1.00}
\newcommand{\aMultiLongPrefillOursVsVllmMax}{1.25}
\newcommand{\aMultiLongPrefillOursVsVllmAvg}{1.12}

\newcommand{\aMultiLongPrefillOursVsNoovlpAvg}{1.21}

\newcommand{\aMultiLongPrefillOursAGRSVsOursARBSOneMin}{0.88}

\newcommand{\aMultiLongPrefillOursAGRSVsOursARBSSixteenMin}{1.12}

%% file: numbers/speedups_1xL40-short.tex
\newcommand{\lSingleShortDecodeOursVsVllmMin}{1.00}
\newcommand{\lSingleShortDecodeOursVsVllmMax}{1.35}
\newcommand{\lSingleShortDecodeOursVsVllmAvg}{1.15}

\newcommand{\lSingleShortDecodeOursVsNoovlpAvg}{1.07}

\newcommand{\lSingleShortDecodeNoovlpVsVllmAvg}{1.07}

%% file: sections/intro-v2.tex
\section{Introduction}

Modern ML frameworks execute models using the \emph{\kernelperop} paradigm~\cite{cheng2026mpk}. In this paradigm, the framework represents an ML model as a computation graph of operators, \eg matrix multiplication, attention, normalization, and collective communication, and runs each operator as a separate GPU kernel. GPU vendors and independent experts implement these kernels. For example, NVIDIA provides matrix multiplication kernels in cuBLAS, attention kernels in cuDNN, and collective communication kernels in NCCL~\cite{cublas,cudnn,nccl}, while experts have developed kernels like FlashAttention~\cite{flashattention}.

\myparab{Modularity of the \kernelperop paradigm.} Since each kernel interacts with the rest of the model only through its input and output tensors, the \kernelperop paradigm provides \emph{composability} and \emph{interoperability} which allow developers to pick and choose for each operator whichever kernel implementation best fits the requirements of their application. The \kernelperop paradigm also provides \emph{debuggability} using tools like Nsight Systems and Nsight Compute~\cite{nsys,ncu} which instrument kernel boundaries. These tools trace kernel launches and analyze each kernel in isolation, allowing developers to attribute a performance anomaly or software bug to a specific kernel.

\myparab{Limitations of the \kernelperop paradigm.}
The \kernelperop paradigm places a \emph{barrier} between dependent kernels since the GPU runtime enforces dependencies at the granularity of entire kernels, either by executing kernels on the same CUDA stream in order or through GPU events between kernels on different streams. Consequently, a kernel cannot start until all thread blocks, or cooperative thread arrays (CTAs), of the kernels it depends on have finished, even when a part of its input is already available. This barrier causes three inefficiencies~\cite{cheng2026mpk,flashmoe}. First, when the last wave of a kernel's CTAs does not fill every streaming multiprocessor (SM), the remaining SMs stay idle until this wave finishes. Second, a communication kernel cannot send any output of the preceding computation kernel until that kernel finishes, which serializes computation and communication. Finally, the next layer's kernel cannot load its weights while the current layer's kernel is running, which prevents the cross-kernel prefetching that weight-bound decode workloads rely on.

\myparab{Megakernels trade modularity for performance.}
Recent work eliminates kernel barriers by fusing the operators of an entire layer or an entire model into a single \emph{megakernel}. MPK and ThunderKittens megakernels fuse the forward pass of an entire model into one kernel~\cite{cheng2026mpk,megatklat}, while FlashMoE fuses the expert computation and inter-GPU communication of a Mixture-of-Experts layer~\cite{flashmoe}. Inside a megakernel, a software scheduler assigns the tiles of all fused operators to SMs and manages the dependencies between individual tiles. This fine-grained scheduling lets the computation, communication, and weight loading of different operators overlap As a result, megakernels outperform the \kernelperop paradigm. However, fusing operators gives up modularity \ie composability, interoperability, and debuggability, of the \kernelperop paradigm. Changing a single operator in a megakernel requires modifying the megakernel itself, which makes it costly to experiment with different kernel implementations or parallelization strategies. Reusing an existing kernel requires rewriting it in the format that the megakernel expects, and closed-source kernels from vendor libraries cannot be fused at all. Since the entire layer or model executes as one kernel, profiling and debugging tools that instrument kernel boundaries fail to attribute a performance anomaly or a failure to an operator.

\myparab{Must we choose between performance and modularity?}
In this work, we ask whether a \kernelperop ML framework can perform on par with megakernels. Kernel barriers are needed only because the GPU runtime enforces dependencies at the granularity of entire kernels. We posit that dependent kernels can instead synchronize on individual tiles of the tensor that one kernel produces and the other consumes. This will enable a kernel to start as soon as the tiles it needs are ready, even while the kernel producing the remaining tiles is still running. 

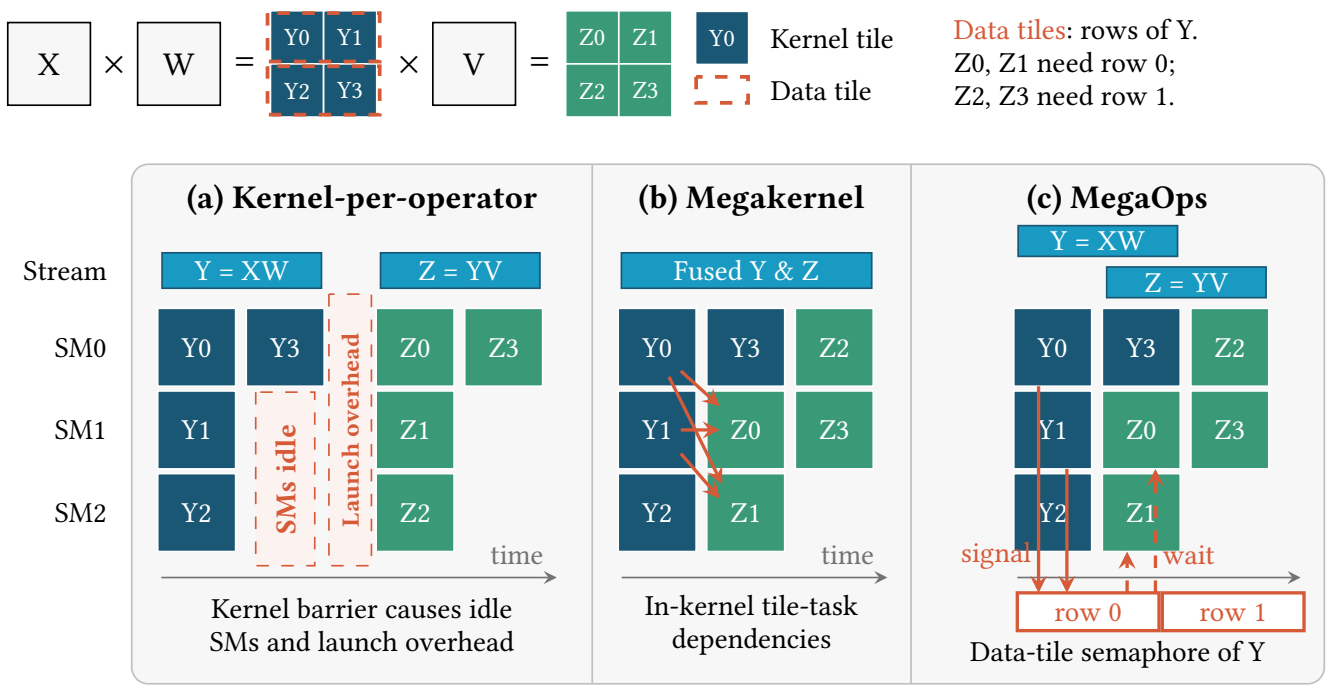
\begin{figure}[t]
  \centering
  \begin{adjustbox}{max width=0.49\textwidth}
  \input{diagram/overview}
  \end{adjustbox}
  \caption{Comparison of (a) \kernelperop, (b) megakernels and (c) \sysname on back-to-back matmuls Y=XW, Z=YV.
    In (c) kernels are unfused and launched on independent streams to overlap. Y is partitioned into row \emph{data tiles}; producer CTAs signal and consumer CTAs wait on semaphores before accessing each data tile.}
  \label{fig:overview}
\end{figure}

\myparab{Data-centric kernel overlap with \sysname.}
Based on this insight, we develop \sysname, a framework that runs each operator as a separate kernel and removes the barriers between dependent kernels. The \sysname framework launches dependent kernels on separate CUDA streams such that the GPU runtime no longer orders the kernels. Instead, \sysname enforces dependencies on individual tiles of the intermediate tensor that the producer kernel writes and the consumer kernel reads. To enable this synchronization, the user passes each kernel its semaphores and a function that maps the kernel's own tiles to data tiles. Kernel developers adapt an existing kernel to \sysname by adding a few lines of code that wait on or signal a semaphore at each tile access.

\myparab{Example.}
Figure~\ref{fig:overview} shows this design on two back-to-back matrix multiplications, $Y = XW$ followed by $Z = YV$. In the \kernelperop paradigm (Figure~\ref{fig:overview}(a)), tile $Y_3$ runs alone in the second wave of the first kernel, and the kernel computing $Z$ cannot use the two idle SMs until $Y_3$ finishes. A megakernel fills these SMs by fusing both matrix multiplications (Figure~\ref{fig:overview}(b)), whereas the \sysname framework achieves the same schedule with two separate kernels (Figure~\ref{fig:overview}(c)). The user partitions $Y$ into two \emph{data tiles}, one per row, and allocates a semaphore for each data tile. The CTAs computing $Y_0$ and $Y_1$ increment the semaphore of row 0 after writing their tiles, and the CTAs computing $Z_0$ and $Z_1$ start on the idle SMs as soon as this semaphore indicates that row 0 is complete, while $Y_3$ is still running. 

\myparab{Data-centric \vs CTA-centric synchronization.}
Synchronizing on data tiles is the key idea that allows \sysname to compose operator kernels. CuSync and FlashOverlap also overlap dependent kernels on separate streams, but they synchronize on the CTAs of the producer kernel instead of the data it produces~\cite{cusync,flashoverlap}. In Figure~\ref{fig:overview}, a CuSync consumer computing $Z_0$ must know which producer CTAs computed row 0 of $Y$. FlashOverlap keeps track of tiles in the order the producer computes them and supports only a matrix multiplication followed by a collective communication operation. In contrast, a consumer in \sysname waits on row 0 of $Y$ regardless of how many producer CTAs computed it. This allows users to replace one producer kernel with a kernel that tiles $Y$ differently, \eg a Stream-K kernel~\cite{osama2023stream}, without modifying the consumer.

\myparab{Results.}
We implement \sysname in Triton, reusing open-source kernels from TritonDistributed~\cite{tritondist} and vLLM~\cite{vllm} and closed-source kernels from cuBLAS and cuDNN, and evaluate it on dense Llama and Qwen3 models with 0.6B to 70B parameters on single H100, A100, and L40 GPUs and on servers with four A100 and eight H100 GPUs. Our \sysname framework achieves 0.89$\times$--1.15$\times$ the throughput of ThunderKittens megakernels on the workloads they are hand-tuned for. Compared with the best-performing baseline in each setting among vLLM, TritonDistributed, and four megakernel systems, 
we achieve on average 1.10$\times$ (1.09$\times$) the throughput in decode and 0.98$\times$ (1.10$\times$) in prefill in single-GPU (multi-GPU) settings.
Since kernels in \sysname compose through data tiles, we can support different parallel schemes that share the same single-GPU compute code path, and require only 2.09$\times$ additional lines of code (LoC) relative to a PyTorch base implementation.

%% file: diagram/overview.tex
  \definecolor{ovtile}{HTML}{195774}%
  \definecolor{ovstream}{HTML}{199AC2}%
  \definecolor{ovaccent}{HTML}{D65A38}%
  \definecolor{ovztile}{HTML}{3A9A78}%
  \begin{tikzpicture}[
    x=1cm, y=1cm,
    font=\rmfamily\small,
    every node/.style={font=\rmfamily\small},
    tile/.style={draw=white, fill=ovtile, minimum size=0.66cm,
      inner sep=0pt, font=\rmfamily\footnotesize, text=white},
    ytile/.style={tile},
    ztile/.style={tile, fill=ovztile},
    mtile/.style={draw=white, fill=ovtile, minimum size=0.45cm,
      inner sep=0pt, font=\rmfamily\scriptsize, text=white},
    mat/.style={draw, fill=black!3, minimum size=0.7cm},
    stream/.style={draw=ovtile, fill=ovstream, text=white, minimum height=0.3cm,
      inner sep=0pt, font=\rmfamily\footnotesize, anchor=west},
    rowlabel/.style={font=\rmfamily\footnotesize, anchor=east},
    title/.style={font=\rmfamily\small\bfseries},
    dep/.style={->, >=stealth, draw=ovaccent, line width=0.8pt},
    taxis/.style={->, >=stealth, black!55},
    sema/.style={draw=ovaccent, line width=0.8pt, fill=white, minimum width=1.2cm,
      minimum height=0.32cm, inner sep=0pt, font=\rmfamily\footnotesize, text=ovaccent},
  ]

    \begin{scope}[shift={(-0.95,2.4)}]
      \node[mat] at (0,0) {X};
      \node at (0.55,0) {$\times$};
      \node[mat] at (1.1,0) {W};
      \node at (1.65,0) {$=$};
      \foreach \i/\c/\r in {0/0/0,1/1/0,2/0/1,3/1/1}
        \node[mtile, fill=ovtile] at (2.1+0.45*\c,0.225-0.45*\r) {Y\i};
      \foreach \r in {0,1}
        \draw[ovaccent, line width=0.9pt, dashed] (1.85,0.43-0.45*\r) rectangle (2.8,0.02-0.45*\r);
      \node at (3.05,0) {$\times$};
      \node[mat] at (3.6,0) {V};
      \node at (4.15,0) {$=$};
      \foreach \i/\c/\r in {0/0/0,1/1/0,2/0/1,3/1/1}
        \node[mtile, fill=ovztile] at (4.6+0.45*\c,0.225-0.45*\r) {Z\i};
      \node[mtile] at (5.7,0.22) {Y0};
      \node[anchor=west, font=\rmfamily\footnotesize] at (6.0,0.22) {Kernel tile};
      \draw[ovaccent, line width=0.9pt, dashed] (5.48,-0.11) rectangle (5.92,-0.33);
      \node[anchor=west, font=\rmfamily\footnotesize] at (6.0,-0.22) {Data tile};
      \node[anchor=west, font=\rmfamily\footnotesize, align=left] at (7.55,0)
        {\textcolor{ovaccent}{Data tiles}: rows of Y.\\
         Z0, Z1 need row 0;\\ Z2, Z3 need row 1.};
    \end{scope}

    \draw[draw=black!25, fill=black!3, rounded corners=4pt] (-0.25,1.55) rectangle (9.85,-2.85);
    \draw[black!25] (3.65,1.55) -- (3.65,-2.85);
    \draw[black!25] (6.35,1.55) -- (6.35,-2.85);
    \node[title] at (1.7,1.25) {(a) Kernel-per-operator};
    \node[title] at (5.0,1.25) {(b) Megakernel};
    \node[title] at (8.1,1.25) {(c) \sysname};

    \begin{scope}
      \node[rowlabel] at (-0.35,0.65) {Stream};
      \node[stream, minimum width=1.35cm] at (0,0.65) {Y = XW};
      \node[stream, minimum width=1.35cm] at (1.85,0.65) {Z = YV};
      \foreach \r in {0,1,2} \node[rowlabel] at (-0.35,-\r*0.7) {SM\r};
      \draw[taxis] (0,-1.95) -- (3.35,-1.95) node[above left, font=\rmfamily\footnotesize, text=black!55] {time};
      \node[ytile] at (0.3,0)    {Y0};
      \node[ytile] at (0.3,-0.7) {Y1};
      \node[ytile] at (0.3,-1.4) {Y2};
      \node[ytile] at (1.05,0)   {Y3};
      \node[ztile] at (2.15,0)    {Z0};
      \node[ztile] at (2.15,-0.7) {Z1};
      \node[ztile] at (2.15,-1.4) {Z2};
      \node[ztile] at (2.9,0)     {Z3};
      \draw[ovaccent, dashed, fill=ovaccent!8] (0.8,-0.38) rectangle (1.3,-1.85);
      \node[text=ovaccent, font=\rmfamily\footnotesize\bfseries, rotate=90, inner sep=0pt] at (1.05,-1.115) {SMs idle};
      \draw[ovaccent, dashed, fill=ovaccent!8] (1.42,0.45) rectangle (1.78,-1.85);
      \node[text=ovaccent, font=\rmfamily\scriptsize\bfseries, rotate=90, inner sep=0pt] at (1.6,-0.7) {Launch overhead};
      \node[text=black, font=\rmfamily\footnotesize, align=center] at (1.7,-2.35)
        {Kernel barrier causes idle\\SMs and launch overhead};
    \end{scope}

    \begin{scope}[shift={(3.9,0)}]
      \node[stream, minimum width=2.12cm] at (-0.01,0.65) {Fused Y \& Z};
      \draw[taxis] (0,-1.95) -- (2.25,-1.95) node[above left, font=\rmfamily\footnotesize, text=black!55] {time};
      \node[ytile] (my0) at (0.3,0)    {Y0};
      \node[ytile] (my1) at (0.3,-0.7) {Y1};
      \node[ytile] (my2) at (0.3,-1.4) {Y2};
      \node[ytile] (my3) at (1.05,0)   {Y3};
      \node[ztile] (mz0) at (1.05,-0.7) {Z0};
      \node[ztile] (mz1) at (1.05,-1.4) {Z1};
      \node[ztile] (mz2) at (1.8,0)     {Z2};
      \node[ztile] (mz3) at (1.8,-0.7)  {Z3};
      \draw[dep] ([shift={(0.2,-0.2)}]my0.center) -- ([shift={(-0.22,0.2)}]mz0.center);
      \draw[dep] ([shift={(0.2,0)}]my1.center) -- ([shift={(-0.22,0)}]mz0.center);
      \draw[dep] ([shift={(0.1,-0.25)}]my0.center) -- ([shift={(-0.2,0.24)}]mz1.center);
      \draw[dep] ([shift={(0.2,-0.2)}]my1.center) -- ([shift={(-0.22,0.12)}]mz1.center);
      \node[text=black, font=\rmfamily\footnotesize, align=center] at (1.1,-2.35)
        {In-kernel tile-task\\dependencies};
    \end{scope}

    \begin{scope}[shift={(7.25,0)}]
      \node[stream, minimum width=1.35cm, minimum height=0.26cm] at (0,0.9) {Y = XW};
      \node[stream, minimum width=1.35cm, minimum height=0.26cm] at (0.75,0.55) {Z = YV};
      \draw[taxis] (0,-1.95) -- (2.4,-1.95);
      \node[ytile] (sy0) at (0.3,0)    {Y0};
      \node[ytile] (sy1) at (0.3,-0.7) {Y1};
      \node[ytile] (sy2) at (0.3,-1.4) {Y2};
      \node[ytile] (sy3) at (1.05,0)   {Y3};
      \node[ztile] (sz0) at (1.05,-0.7) {Z0};
      \node[ztile] (sz1) at (1.05,-1.4) {Z1};
      \node[ztile] (sz2) at (1.8,0)     {Z2};
      \node[ztile] (sz3) at (1.8,-0.7)  {Z3};
      \node[sema] (s0) at (0.6,-2.24) {row 0};
      \node[sema, anchor=west] (s1) at (s0.east) {row 1};
      \node[font=\rmfamily\footnotesize, text=black] at (0.85,-2.6)
        {Data-tile semaphore of Y};
      \draw[dep] (0.18,-0.33) -- (0.18,-2.08);
      \draw[dep] (0.42,-1.03) -- (0.42,-2.08);
      \draw[dep, dashed] (0.93,-2.08) -- (0.93,-1.73);
      \draw[dep, dashed] (1.17,-2.08) -- (1.17,-1.03);
      \node[font=\rmfamily\footnotesize, text=ovaccent, inner sep=1pt, anchor=east] at (0.16,-1.76) {signal};
      \node[font=\rmfamily\footnotesize, text=ovaccent, inner sep=1pt, anchor=west] at (1.19,-1.76) {wait};
    \end{scope}
  \end{tikzpicture}%

%% file: sections/method.tex
\section{\sysname Framework}
\label{sec:method}

The \sysname framework consists of two components, a data-tile kernel interface that lets dependent kernels synchronize on tiles of the tensors between them (\S\ref{subsec:data-tile}, \S\ref{subsec:kernel-interface}), and a runtime called AsyncStream that launches kernels on multiple CUDA streams so that the GPU runtime does not order them (\S\ref{subsec:asyncstream}). By the \emph{interface} of a kernel, we mean the arguments through which a caller launches it. In the traditional \kernelperop paradigm, this interface consists of the kernel's input and output tensors and a few configuration parameters. This interface does not reveal how the kernel divides its work among CTAs. \sysname extends this interface with arguments that tell a kernel when each part of its input is ready and where to record that each part of its output has been written. We use the back-to-back matrix multiplications in Figure~\ref{fig:overview} as a running example.

\subsection{A Data-centric Principle for Kernel Composition}
\label{subsec:data-tile}
In the \kernelperop paradigm, the interface of a kernel consists of its input and output tensors. Since a caller interacts with a kernel only through these tensors, it does not depend on how the kernel is implemented or tiled. As a result, one kernel can replace another without changes elsewhere in the model. Overlapping dependent kernels, as proposed by \sysname, appears to break this composability. When a producer and a consumer kernel run concurrently on separate streams (Figure~\ref{fig:overview}(c)), a consumer CTA must wait until the producer CTAs that write its input have finished. If the consumer waits on producer CTAs, it must know how the producer assigns tiles to CTAs, which couples the tilings of the two kernels. \sysname avoids this coupling by synchronizing on data instead of CTAs.

\myparab{Data tile and readiness semaphore.}
\sysname asks the user to partition the tensor that connects the two kernels into \emph{data tiles}, independent of the tiling of either kernel. The consumer waits until data tiles it reads are ready. Each data tile is assigned a multi-dimensional coordinate. The user creates a \emph{semaphore} for each data tile to track if the producer has fully written it. Inspired by \ETC~\cite{etc}, semaphores are ordinary integer tensors owned and managed by users, so that semaphore hangs can be debugged on the user side by inspecting their values with a timeout mechanism we provide. Users then adapt both the producer and consumer kernels to the data tiling via our kernel interface, so that a producer CTA updates the semaphore to signal that a data tile is ready, and a consumer CTA checks the semaphore before it reads the data tile. As a result, dependent kernels compose freely as long as they are adapted to the same data tiling.

\subsection{Kernel Interface}
\label{subsec:kernel-interface}

\begin{lstlisting}[caption=\sysname kernel interface,float=t,tabsize=2,label=lst:interface]
def kernel(...,
 in_semaphore: Tensor # data tile semaphore 
                      # for input tensor
 in_map: Callable # kernel-tile-->data-tile
                  # N-D coordinate mapping
 in_target: int # target semaphore value
                # for readiness
 in_sema_extent: tuple # N-D data-tile 
                       # coordinate extent
 # similar arguments for output
 out_semaphore: Tensor
 out_map: Callable
 out_sema_extent: tuple
 # optional tile size configurations
 tile_config: dict
 # optional remap to reschedule CTAs
 # in a different order
 cta_remap: Callable): ...
\end{lstlisting}
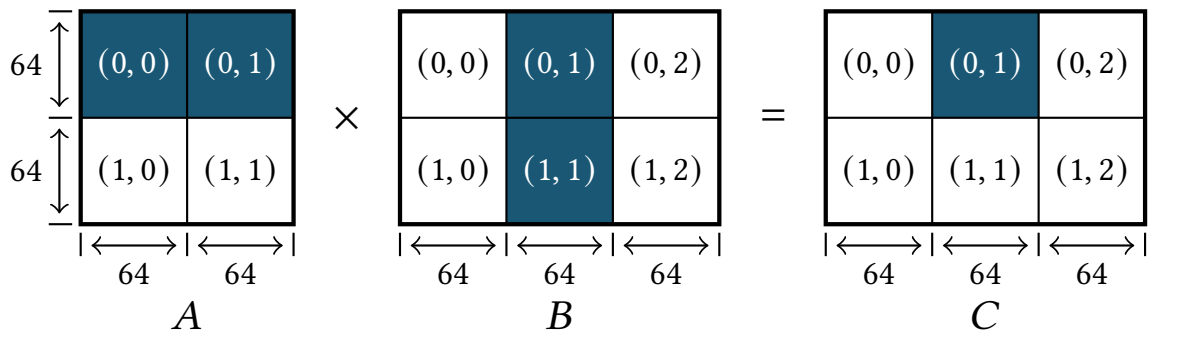
\begin{figure}[t]
  \centering
  \definecolor{tileteal}{HTML}{195774}
  \resizebox{0.9\linewidth}{!}{%
  \begin{tikzpicture}[
    x=0.7cm, y=0.7cm,
    every node/.style={font=\small},
    tile label/.style={font=\scriptsize},
    selected/.style={fill=tileteal}
  ]
    \fill[selected] (0,1) rectangle (2,2);
    \fill[selected] (4,0) rectangle (5,2);
    \fill[selected] (8,1) rectangle (9,2);

    \foreach \origin/\name/\cols in {0/A/2,3/B/3,7/C/3} {
      \draw[thick] (\origin,0) rectangle (\origin+\cols,2);
      \pgfmathtruncatemacro{\lastcol}{\cols-1}
      \foreach \boundary in {1,...,\lastcol} {
        \draw (\origin+\boundary,0) -- (\origin+\boundary,2);
      }
      \draw (\origin,1) -- (\origin+\cols,1);
      \foreach \boundary in {0,...,\cols} {
        \draw (\origin+\boundary,-0.08) -- (\origin+\boundary,-0.3);
      }
      \foreach \col in {0,...,\lastcol} {
        \draw[<->, shorten <=2pt, shorten >=2pt] (\origin+\col,-0.2) -- (\origin+\col+1,-0.2)
          node[midway, below, font=\scriptsize] {$64$};
      }
      \node[inner sep=0pt, outer sep=0pt] at (\origin+\cols/2,-0.85) {$\name$};
      \foreach \row in {0,1} {
        \foreach \col in {0,...,\lastcol} {
          \def\labelcolor{black}
          \ifnum\origin=0 \ifnum\row=0 \def\labelcolor{white}\fi\fi
          \ifnum\origin=3 \ifnum\col=1 \def\labelcolor{white}\fi\fi
          \ifnum\origin=7 \ifnum\row=0 \ifnum\col=1
            \def\labelcolor{white}\fi\fi\fi
          \node[tile label, text=\labelcolor] at (\origin+\col+0.5,1.5-\row)
            {$(\row,\col)$};
        }
      }
    }
    \node at (2.5,1) {$\times$};
    \node at (6.5,1) {$=$};
    \foreach \boundary in {0,1,2} {
      \draw (-0.08,\boundary) -- (-0.3,\boundary);
    }
    \foreach \row in {0,1} {
      \draw[<->, shorten <=2pt, shorten >=2pt] (-0.2,\row) -- (-0.2,\row+1)
        node[midway, left, font=\scriptsize] {$64$};
    }
  \end{tikzpicture}%
  }
  \caption{Kernel tiles and coordinates for $C=A B$ with tile sizes of 64 on row, column, and reduction dimensions.
    The blue kernel tiles are accessed when computing $C$ tile $(0,1)$.
    }
  \label{fig:matmul-kernel-tiling}
\end{figure}

To synchronize on data tiles, a kernel needs to know how the user has partitioned each tensor into data tiles. The kernel must wait on or signal the corresponding semaphores before it reads or after it writes a data tile. Our kernel interface, summarized in Listing~\ref{lst:interface}, lets users pass this information to the kernel. We currently support synchronization on one input tensor and one output tensor, so a kernel accepts a single \texttt{in\_semaphore} and a single \texttt{out\_semaphore}, which contain the data-tile semaphores of these tensors.

Kernels typically partition the output tensor into tiles and assign each tile to a CTA, 
so each CTA accesses a tile of every input and output tensor, which we call a \emph{kernel tile}. We require kernels to expose their tiling scheme and assign each kernel tile a multi-dimensional coordinate in natural language as part of the function documentation, so that the user knows which part of the data each kernel tile accesses. For example, Fig.~\ref{fig:matmul-kernel-tiling} shows the kernel tiling and coordinates in a matmul kernel with $64$ as tile sizes in all dimensions. The kernel developer can also choose to expose coarser-grained kernel tiles than the actual kernel implementation to reduce synchronization overhead, like merging tiles in the same row into a bigger tile.

Using this documentation, the user passes mappings and extents that translate kernel tiles into data tiles. The mapping functions, \texttt{in\_map} and \texttt{out\_map}, map a kernel-tile coordinate to the coordinate of the first data tile it intersects, and the extents, \texttt{in\_sema\_extent} and \texttt{out\_sema\_extent}, specify how many data tiles the kernel tile spans in each dimension. Together, they identify the semaphore entries that a CTA synchronizes on. Our kernel-side contract is that whenever the kernel accesses a kernel tile, it uses the mapping and extent to find the corresponding data tiles, and then waits on \texttt{in\_semaphore} until each entry reaches \texttt{in\_target}, or signals \texttt{out\_semaphore}. Here, \texttt{in\_target} specifies the number of kernel tiles needed to fully produce the data tile, which can be $\geq1$ as a data tile may likewise have multiple intersecting kernel tiles.

Finally, the user can pass a \texttt{tile\_config} to customize the kernel's tile sizes, akin to the block sizes in many Triton kernels. %
Optionally, kernels can expose a \texttt{cta\_remap} interface that remaps CTA IDs to let users change the CTA processing order, \eg to follow the order in which the producer makes data tiles ready (\S\ref{subsec:reorder-prefetch}).

\begin{lstlisting}[caption={Back-to-back matmul of $Y=XW, Z=YV$ in \sysname. \texttt{cdiv} is ceiling division. We assume either data tile size divides the kernel tile sizes or the reverse for simplicity. We discuss indivisible cases later.},float=t,tabsize=2,morekeywords={Callable,Tensor,zeros,matmul,tune\_config,cdiv,signals\_per\_ktile},label=lst:ex]
m, hidden_size = x.shape

# pick rowwise plain data tiling scheme
dtile_m = 64 # data tile size
y_sema = zeros(cdiv(m, dtile_m),
  dtype=torch.int32, device=x.device)

# adapt producer
t_cfg = tune_config('matmul', x, W)
ktile_m = t_cfg['tile_m'] # kernel tile size
y = matmul(x, W, t_cfg,
  out_semaphore=y_sema,
  out_map=f'lambda ktile_coord:ktile_coord[0]
                 * {ktile_m} // {dtile_m}',
  out_sema_extent=cdiv(ktile_m, dtile_m))
# number of ktiles per dtile
y_target = cdiv(y.shape[1], t_cfg['tile_n'])
  * cdiv(dtile_m, ktile_m)

# adapt consumer
t_cfg = tune_config('matmul', y, V)
ktile_m = t_cfg['tile_m']
z = matmul(y, V, t_cfg,
  in_semaphore=y_sema,
  in_map=f'lambda ktile_coord:ktile_coord[0]
                * {ktile_m} // {dtile_m}',
  in_sema_extent=cdiv(ktile_m, dtile_m)),
  in_target=y_target)
\end{lstlisting}

\begin{figure}[t]
  \centering
  \definecolor{tileteal}{HTML}{195774}
  \definecolor{ovaccent}{HTML}{D65A38}
  \resizebox{0.8\linewidth}{!}{%
  \begin{tikzpicture}[
    x=0.85cm, y=0.85cm,
    every node/.style={font=\small},
    tile label/.style={font=\scriptsize},
    selected/.style={fill=tileteal}
  ]
    \foreach \origin/\name/\cols in {7/y/3} {
      \draw[thick] (\origin,0) rectangle (\origin+\cols,2);
      \pgfmathtruncatemacro{\lastcol}{\cols-1}
      \foreach \boundary in {1,...,\lastcol} {
        \draw (\origin+\boundary,0) -- (\origin+\boundary,2);
      }
      \draw (\origin,1) -- (\origin+\cols,1);
      \foreach \boundary in {0,...,\cols} {
        \draw (\origin+\boundary,-0.18) -- (\origin+\boundary,-0.4);
      }
      \foreach \col in {0,...,\lastcol} {
        \draw[<->, shorten <=2pt, shorten >=2pt] (\origin+\col,-0.3) -- (\origin+\col+1,-0.3)
          node[midway, below, font=\scriptsize] {$128$};
      }
      \ifnum\origin=7
        \foreach \d/\lo/\hi in {0/1.53/2.07, 1/1.03/1.47, 2/0.53/0.97, 3/-0.07/0.47} {
          \draw[dashed, ovaccent, line width=0.9pt]
            (6.9,\lo) rectangle (10.1,\hi);
        }
      \fi
      \ifnum\origin=7
      \foreach \row in {0,1} {
        \foreach \col in {0,...,\lastcol} {
          \def\labelcolor{black}
          \def\labelbg{white}
          \node[tile label, text=\labelcolor, inner sep=1pt] at (\origin+\col+0.5,1.5-\row)
            {$(\row,\col)$};
        }
      }
      \fi
    }

    \foreach \boundary in {0,1,2} {
      \draw (6.58,\boundary) -- (6.8,\boundary);
    }
    \foreach \row in {0,1} {
      \draw[<->, shorten <=2pt, shorten >=2pt]
        (6.68,\row) -- (6.68,\row+1)
        node[midway, left, font=\scriptsize] {$128$};
    }

    \foreach \d in {0,1,2,3} {
      \node[anchor=west, tile label, text=ovaccent, inner sep=0pt] at (10.12,1.75-0.5*\d) {$(\d,)$};
    }

    \draw[thick] (11.1,1.05) rectangle (11.9,1.55);
    \node[tile label] at (11.5,1.3) {$(0,0)$};
    \node[anchor=west, font=\scriptsize] at (12.0,1.3) {Kernel tile};
    \draw[dashed, ovaccent, line width=0.9pt] (11.1,0.4) rectangle (11.9,0.85);
    \node[tile label, text=ovaccent] at (11.5,0.625) {$(0,)$};
    \node[anchor=west, font=\scriptsize] at (12.0,0.625) {Data tile};
  \end{tikzpicture}%
  }
  \caption{Mapping kernel tiles of $Y=XW$ to row-wise data tiles in Listing~\ref{lst:ex}.
    Each kernel tile intersects two data tiles, \eg kernel tile $(1,1)$ maps to data tiles 2 and 3, encoded by \texttt{out\_map([1,1])}
    $=1\cdot128\mathbin{//}64=2$ with extent of 2. Each data tile receives
    \texttt{y\_target}$=3$ signals.}
    
  \label{fig:ex-out-mapping}
\end{figure}
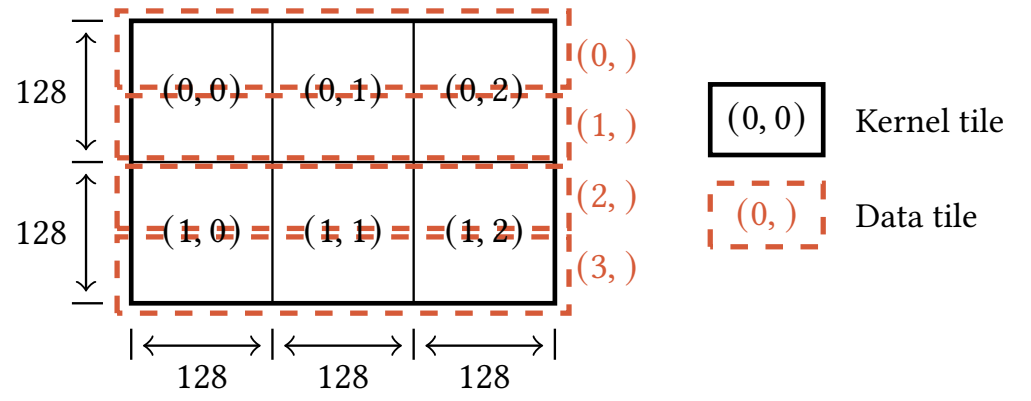

\myparab{Example.} Listing~\ref{lst:ex} shows how our interface can be used to implement a back-to-back matmul $Y=XW, Z=YV$.
First, the user decides the data tiling scheme and tile size of the connecting intermediate tensor $Y$, which uses row-wise tiling with size 64.
Next, to run the first matmul, the user tunes the tile size with \texttt{tune\_config} and gets the kernel tile sizes.
For simplicity, the example assumes the tuned tile size can divide or be divided by 64.
To adapt the producer to $Y$'s data tiling, the user maps the kernel tile to $Y$'s data tile, where the kernel has tile size \texttt{ktile\_m} and $Y$ has \texttt{dtile\_m}. \texttt{ktile\_coord[0] * ktile\_m // dtile\_m} maps the coordinate of the first element, which is passed to \texttt{out\_map}.
To complete the mapping, the user computes the number of data tiles a kernel tile covers with \texttt{cdiv(ktile\_m, dtile\_m)} and passes it to \texttt{out\_sema\_extent}. The user computes the number of signals every data tile will receive and stores it in \texttt{y\_target} for the consumer. Fig.~\ref{fig:ex-out-mapping} shows the mapping.
The user adapts the consumer kernel similarly on the input side.

\begin{lstlisting}[caption={Kernel-side matmul integration example. Starting from an ordinary Triton GeMM (tile addressing omitted), the developer adds the \texttt{sync\_ctx} argument and the lines marked \texttt{\#+}: L4 signals the start of the kernel; L8 reorders CTA IDs; L12 waits with $A$'s kernel tile coordinate; L19 signals with $C$'s kernel tile coordinate.},float=t,tabsize=2,morekeywords={_wait_input_cta,_signal_output_cta,_cta_remap,_kernel_begin_cta},label=lst:kernel-integration]
@triton.jit
def matmul_kernel(A, B, C, M, N, K, 
    TILE_M, TILE_N, TILE_K, sync_ctx):
  _kernel_begin_cta(sync_ctx)            #+
  pid = tl.program_id(0)
  nn = tl.cdiv(N, TILE_N)
  _pids = (pid // nn, pid %
  _pids = _cta_remap(sync_ctx, _pids)    #+
  pid_m, pid_n = _pids                   #+
  _ktile_in = (pid_m,) #+ expose row-tiling
  _ktile_out = (pid_m, pid_n)#+ grid-tiling
  _wait_input_cta(sync_ctx, _ktile_in)   #+
  acc = tl.zeros((TILE_M, TILE_N), ...)
  for k in range(0, K, TILE_K):
    a = tl.load(...)
    b = tl.load(...)
    acc = tl.dot(a, b, acc)
  tl.store(..., acc)
  _signal_output_cta(sync_ctx, _ktile_out)#+
\end{lstlisting}

\ignore{
\begin{lstlisting}[language=Python,caption={\jk{drop or go to appendix} Host-side matmul integration example. The developer extends the ordinary launcher \texttt{matmul} for the kernel in Listing~\ref{lst:kernel-integration} only to forward \texttt{sync\_ctx}, wraps it with \texttt{with\_semaphore\_wrapper} to expose the \sysname interface, and registers it with \texttt{register\_megaops} together with \texttt{signals\_per\_ktile} (one signal per $C$ kernel tile, since each is written by a single CTA) and the tile-size tuning functions.},float=t,tabsize=2,morekeywords={register_megaops,with_semaphore_wrapper},label=lst:kernel-integration-host]
def matmul(x, W, tile_config, sync_ctx):
  y = torch.empty(...)
  grid = ...  # number of C tiles
  matmul_kernel[grid](x, W, y, ..., sync_ctx)
  return y

matmul = with_semaphore_wrapper(matmul)
def signals_per_ktile(cfg):
  return 1
def tune_config_heuristic(x, W):
  return dict(tile_m=128, tile_n=128, tile_k=32)
def tune_config_candidates(x, W):
  return [dict(tile_m=bm, tile_n=bn, tile_k=bk)
    for bm in (64, 128) for bn in (64, 128)
    for bk in (32, 64)]
register_megaops(matmul,
  signals_per_ktile,
  tune_config_heuristic,
  tune_config_candidates)
\end{lstlisting}
}

\myparab{Kernel integration.}
Integrating a kernel to support our interface is straightforward and follows programming patterns common in many of today's kernel design conventions such as prologue, epilogue, and tile scheduler from CUTLASS~\cite{cutlass}.
We provide utility functions to facilitate the integration, including \texttt{\_wait\_input\_cta} for input and \texttt{\_signal\_output\_cta} for output semaphore synchronization, and the kernel only needs to invoke them with the kernel tile coordinates. The utility functions convert them to data tile ones and perform the corresponding synchronization under the hood.
Our prototype is based on Triton, and Listing~\ref{lst:kernel-integration} shows the integration of a GeMM kernel as an example. It only requires adding a few lines of code, demonstrating that the effort is small to integrate a kernel in \sysname.

\myparab{Handling irregular and indivisible tile mapping.} We further extend our interface to accept a callable for \texttt{in/out\_sema\_extent} and \texttt{in\_target}, which accept the kernel-tile and data-tile coordinate, respectively. This allows the value to vary from tile to tile by computing it dynamically, similar to the \texttt{in/out\_map}s.
This is useful when different data tiles have different numbers of kernel tiles intersecting, which happens often with indivisible tile sizes. A function-based extent and target computation can address the issue.
To increase flexibility, the user can add any additional arguments at the kernel launch and pass them to these functions. Note that our functions support any arbitrary Triton functions, so users may even precompute these function values and store them in a tensor and load them back when needed. This is useful for highly irregular tiling where no closed-form formula exists for extents or targets.

\myparab{Utility functions to simplify user programming in plain tilings.}
In a \emph{plain tiling}, a tensor is partitioned regularly into consecutive chunks of a fixed size in each dimension. Irregular tilings are rare. Most often a plain tiling scheme is used for both data and kernel tiles. The user programming effort can be further reduced in this case using our utility functions. We observe that the programming pattern is homogeneously matching two tile grids with different grid sizes on the same tensor.
For example, in Fig.~\ref{fig:ex-out-mapping} computing $Y$'s kernel-to-data tile extent can be abstracted as computing how many dashed grid cells a solid grid cell maps to.
We therefore provide the following utility functions using this grid abstraction so that users do not need to calculate the formula manually.
\begin{itemize}
    \item calc\_extent($dtile$, $shape$, $ktile$) returns a lambda function (in a Python string) that accepts a kernel tile coordinate and calculates the extent, \ie the number of data tiles that intersect with it, potentially dynamically. $dtile$ and $ktile$ represent the data and kernel tile sizes, respectively, and $shape$ represents the full shape of the tensor.
    \item calc\_target($dtile$, $shape$, $ktile$) similarly returns a lambda function that accepts a data tile coordinate and returns the number of kernel tiles that intersect.
\end{itemize}
Using them, the calculation formulas for y\_target and extents in Listing~\ref{lst:ex} can both be replaced with a simple function call.

\myparab{Supporting GPUDirect peer-to-peer (P2P) memory access.} Communication kernels, especially in the scale-up domain, often rely on direct P2P memory access to reduce local data copy and achieve efficient communication. 
The kernel typically relies on symmetric memory~\cite{nvshmem,symm}, where all ranks allocate a tensor of the same size and can address peer tensors directly.
To support this pattern, we prepend a rank dimension to our data and kernel tile coordinates to identify remote tiles. %
The semaphores for a symmetric memory tensor will also be stored using symmetric memory to let a kernel synchronize the readiness of a remote data tile.

\myparab{Supporting closed-source kernels.} Triton kernels may not always have the best performance, and other kernels with higher performance may not have source code available for the above integration, such as the cuBLAS and copy engine kernels.
Inspired by prior decomposition techniques~\cite{googletp,asynctp,megatd}, we split the output tensor into a few chunks and launch one smaller kernel, which we call a \emph{subkernel}, for each chunk. 
We conceptually regard these subkernels as a coarse-grained set of CTAs of the undecomposed kernel, and invoke the provided wait and signal functions on the kernel tiles they access at subkernel boundaries.
To make this happen, we run subkernels in two separate streams in a round-robin order to avoid inter-subkernel barriers, and wrap each with a tiny wait and signal synchronization kernel to control when to start a subkernel and to notify completion. We launch a proxy kernel on a third stream to interact with these synchronization kernels, so that we can invoke \texttt{\_wait\_input\_cta} and \texttt{\_signal\_output\_cta} in the proxy before starting a subkernel and after one finishes. %

\subsection{AsyncStream Runtime}
\label{subsec:asyncstream}
We develop a runtime called AsyncStream to overlap the kernels by dispatching to separate streams in a round-robin schedule. The number of streams is configurable, trading between interference and overlap.
Exposed to the outside, AsyncStream provides an abstraction as if all \sysname kernels are run under one single stream, except that the stream imposes no inter-kernel barrier.
AsyncStream then offers common wait and block stream synchronization primitives, so that it can interact with other regular streams and coexist with existing unintegrated kernels,  maximizing compatibility and enabling gradual integration.
In doing so, AsyncStream addresses the following challenges to achieve correct and efficient scheduling and overlap.

\myparab{Deadlock-free scheduling.}
It guarantees that a kernel launched earlier can secure hardware resources before later ones to avoid deadlocks, using the same wait kernel technique from CuSync: a wait kernel is launched on the same stream before every kernel launch to delay the current kernel from being scheduled until after the prior ones have secured their hardware resources. Empirically, we observe that CUDA schedules all CTAs of a kernel before scheduling the next. Thus, we emit a signal from the prior kernel whenever any CTA has started using \texttt{\_kernel\_begin\_cta} (Listing~\ref{lst:kernel-integration}), which the wait kernel waits on.

\myparab{Overlapping communication and computation kernels with spatial partitioning.}
Without spatial partitioning, the earlier kernels, especially computation ones, will often occupy the whole GPU, leaving no resources available to overlap the second kernel.
We address this problem by spatially carving out a small number of SMs for running communication kernels, as a few SMs are sufficient to supply sufficient bandwidth for overlapping~\cite{qiang2026syncopate,megatktput}.
We use Green Contexts~\cite{greenctx}, which can assign streams to a GPU partition, and launch kernels to corresponding streams based on kernel types.

\myparab{Automatic memory management.}
PyTorch's CUDA memory pool assumes a tensor is used on a single stream, so under multi-streaming it may reuse a tensor whose kernel is still running. Its stream-recording workaround incurs host-device synchronization and is incompatible with CUDA graphs. 
Furthermore, PyTorch's memory pool is nondeterministic in the reuse outcome, which causes issues when reusing symmetric memory.
This is because each symmetric memory tensor has its paired peer tensors, and all ranks must make the same decision and reuse the same paired ones. 
We address both automatically with delayed tensor release and a consistent symmetric memory pool.
We cap the number of concurrent kernels at $N$ using CUDA events and hold a reference to each tensor a kernel uses until $N$ more kernels have launched, by which time that kernel must have finished. For symmetric tensors, we additionally hold the reference until all peers are known to have passed the $N$-kernel mark. %
We rely on user annotation of cross-peer data dependency to trace peer progress without extra synchronization (details in Appendix~\ref{app:mem}).
Finally, to guarantee peer consistency, symmetric tensors are allocated from a dedicated pool that reuses only exact-size matches and returns the earliest freed tensor, so that all ranks make the same decision. We also run Python's garbage collection (GC) manually in the same schedule across peers for consistency, and freeze prior generations to reduce the overhead. Appendix~\ref{app:mem} details the design.

\section{Overlap-aware Optimization}
\label{sec:optimization}

Launching dependent kernels on separate streams in \sysname removes kernel barriers, but by itself it does not deliver the performance of megakernels. We find two causes of this behavior. First, existing kernels are written and tuned under the assumption that they own the GPU, so they do not take advantage of the opportunities that overlap creates. For example, a kernel could prefetch its weights while the preceding kernel is still running, and it could process its tiles in the order in which the producer finishes them, but existing kernels do neither (\S\ref{subsec:reorder-prefetch}). Second, overlapping kernels compete for SMs, and the CUDA scheduler decides how to place their CTAs. A poor placement leaves SMs idle or slows down the execution (\S\ref{subsec:occupancy-cap}). We address the first cause by reordering instructions and the second with a mechanism that regulates CTA placement. Since the best settings for both depend on how the kernels in a model overlap, we tune them jointly with a graph-level autotuner (\S\ref{subsec:graph-tune}).

\subsection{Dependency-aware Reordering and Prefetching}
\label{subsec:reorder-prefetch}
The existing \kernelperop paradigm assumes all inputs are ready before kernel launches. Kernels thus often adopt a swizzled CTA schedule to optimize data locality. Under kernel overlap, this assumption breaks: some CTAs that are scheduled early may stall on unavailable inputs while others with input ready are scheduled to start later, creating pipeline bubbles. Inspired by CuSync, we expose a \texttt{cta\_remap} interface (Listing~\ref{lst:interface}\&\ref{lst:kernel-integration}) that allows users to align CTA scheduling with producer data ready order. For example, in an allgather--GeMM overlap, the push-based allgather~\cite{asynctp} produces local shards (via local memcpy) before remote ones. Local GeMM CTAs can run first by remapping $ktile[0]$ to $(ktile[0]+d)\% m$, where $d$ is the first local tile index and $m$ the number of row tiles. In low-batch decoding, GeMM weight loading dominates execution time. Kernel overlap allows us to prefetch weights during the preceding kernel's execution. We lift a tunable number of weight tile loads to the beginning of the kernel code, before semaphore waits, to prefetch. More prefetches hide more latency but require earlier kernel launches (so more streams) and consume more SMEM/registers. Our graph pass autotunes the prefetch depth. %
\subsection{Overlap Regulation with SMEM Adjustment}
\label{subsec:occupancy-cap}

Overlapping kernels interfere with each other in CTA-to-SM scheduling. We observe that spatially sharing the same SM between two overlapping kernels does not always happen, even when the SM has enough resources left. This causes severe performance loss to kernels with a smaller number of waves, such as persistent GeMMs or under small batch sizes: they cannot use the SMs occupied by prior overlapping kernels, and CUDA eagerly schedules all CTAs to the remaining SMs at the moment. When prior kernels finish, the released SMs become idle, wasting resources.

We address it with CTA launch blocking. We observe that prior kernels are often at the end of execution when the next kernel starts launching CTAs. Therefore, we block some launches of the next kernel's CTAs until prior kernels finish, using an SMEM padding technique to force a CTA admission control. %
In detail, as each SM has a limited SMEM capacity, we can increase SMEM of a kernel to limit the number of colocated CTAs. By allocating a large enough SMEM, the unoccupied SM subset no longer has enough SMEM to host all CTAs, which forces CUDA to stop launching new CTAs. 
We design an occupancy limit parameter $c$, the maximum number of CTAs allowed, and set the SMEM allocation to $\frac{C}{c+1}+1$ where $C$ is the GPU SMEM capacity. Our graph pass autotunes $c$. %

On the other hand, when two kernels use complementary hardware resources, such as tensor cores versus CUDA cores, it is beneficial to \emph{reduce} SMEM usage of a kernel to encourage spatial SM sharing.
To do so, we tune the number of stages, a common knob existing in modern kernels, to reduce SMEM usage, and autotune it in our graph pass.

\subsection{Graph-level Autotuning}
\label{subsec:graph-tune}
The parameters mentioned above are sensitive to how kernels overlap and hence cannot be determined on a per-kernel basis. We thus develop a graph-level tuning algorithm that runs all kernels to optimize these parameters. To search efficiently, we adopt an iterative tuning approach where we vary only one parameter and fix all others at a time. %
Each parameter enumerates over a set of predefined candidate values and profiles performance to select the best one. To reduce overhead for transformer models that have homogeneous layers, only the first few layers are used for profiling.
We also tune the number of streams and number of decomposed chunks, as well as other customized knobs required to be tuned in an overlap environment.

%% file: sections/impl.tex
\section{Implementation}
\label{sec:impl}
We prototype our design using Triton and TritonDistributed.
The functions in the interface are implemented as Triton functions, taking advantage of Triton's support for passing Triton functions as kernel arguments. Lambda functions are supported by translating them to a Triton function at runtime.
PyTorch symmetric memory is used to support peer memory and multicast address mapping.
Appendix~\ref{app:impl} lists the kernels we support and details our scheduling optimization of closed-source kernels.
We are compatible with and rely on CUDA graphs to reduce launch overhead.

%% file: sections/eval.tex
\section{Evaluation}

We answer the following questions in the evaluation.
\begin{packeditemize}
    \item Can \sysname match megakernel performance in model serving, in both single- and multi-GPU settings?
    \item Can \sysname match handwritten libraries in computation-communication overlap workloads?
    \item How much is the coding effort on the user side?
    \item Is graph tuning essential to obtain high performance?
\end{packeditemize}
\subsection{Evaluation Setup}
\label{sec:eval-setup}
We use both single-GPU and single-node multi-GPU settings to evaluate our system. Single-GPU settings consist of A100 (40GB SXM), L40, and H100 (80GB HBM3), while multi-GPU settings consist of 4$\times$A100 GPUs fully connected via direct P2P NVLink and 8$\times$H100 connected via NVLS, with 100 GB/s and 450 GB/s unidirectional bandwidth, respectively. %
We describe the multi-GPU results on 8$\times$H100 in this section and defer the 4$\times$A100 results to Appendix~\ref{sec:ext-eval-multi-a100}, which show similar trends. We use CUDA 13.1 and PyTorch 2.11.0 for all experiments.

\myparab{Workloads.}
We evaluate the performance of a single batch of requests serving (\S\ref{subsec:eval:e2e}) using long and short contexts, over a set of batch sizes. The long decode uses 1024 for prompt length and 1024 for output length, while the short decode uses 128 for prompt length and 64 for output length. We use Qwen3-0.6B, 1.7B, 8B, Llama-3.2-1B, and Llama-3.1-70B for this evaluation, which cover a variety of dense models. 
To evaluate the effect of computation-communication overlap more directly, we use allgather-GeMM overlap from a TritonDistributed benchmark (\S\ref{subsec:eval:comm-comp-ovlp}) which runs MLP allgather and up-projection overlap under tensor parallelization. The weight shapes come from Llama-3.1-8B, 70B, and 405B, and Qwen2-72B, GPT3-175B with varying batch sizes, covering both small and large shapes.

\myparab{Baselines.} We compare against \emph{vLLM}, \emph{TritonDistributed}, megakernel compilers and frameworks \emph{MPK}, MegaTritonKernel~\cite{megatd,zheng2026ditron} (referred to as \emph{MegaTD}) from TritonDistributed, and \emph{VDCores}~\cite{vdcores}, and specialized megakernels \emph{ThunderKittens} including both the single-GPU low-latency and the 8-GPU high-throughput versions. We also run a no-overlap version of our system by disabling multi-streaming and removing the prefetch and SMEM padding.
For clarity, we omit systems on prefill or decode plots when they fail to run or perform poorly (Appendix~\ref{appendix:eval-setup-details}).
For communication-computation overlap, we compare against \emph{AsyncTP}~\cite{asynctp} and TritonDistributed.
We run each system 5$\times$ and report the median normalized against vLLM. We report the speedup against the best other systems on the bar plots.

\begin{figure*}[t]
\centering
    \includegraphics[width=0.99\linewidth]{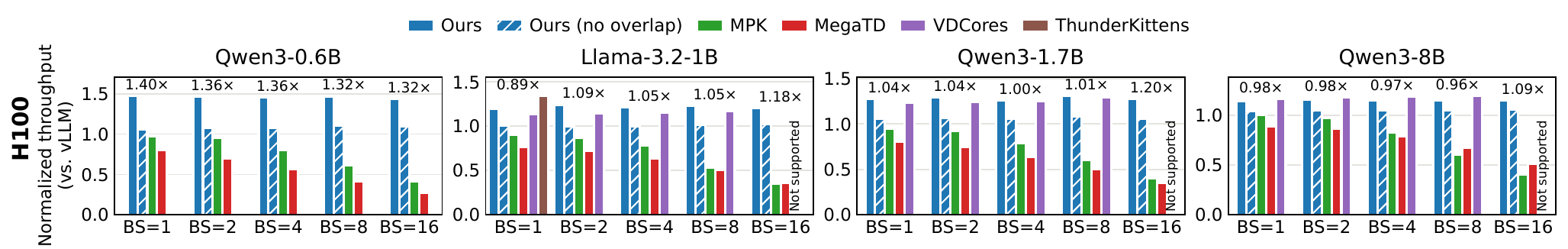}
    \caption{Short decode throughput on H100 using 128 prompt tokens and 64 output tokens. Numbers above bars indicate the speedup of ``Ours'' over the best of the other systems. A100 and L40 results are in Fig.~\ref{fig:short-decode-a100-l40} in Appendix.}
    \label{fig:short-decode}
\end{figure*}
\myparab{\sysname implementation and parallel schemes.}
We implement the dense Llama and Qwen3 model families using our layer-level interface. For multi-GPU execution, we implement two parallel schemes: an allreduce (AR) scheme from the traditional Tensor Parallelism~\cite{shoeybi2019megatron}
and an allgather-reducescatter (AG-RS) scheme from \TK megakernels~\cite{megatktput}.
AR is used for decode, which has a smaller batch. %
For prefill, we manually profile both schemes to pick a batch size threshold: AG-RS is used from BS $\geq 8$ for Qwen3-8B and BS $\geq 2$ for Llama-3.1-70B on 8$\times$H100, and from BS $\geq 2$ on 4$\times$A100.
Most intermediate tensors use a row-wise data tiling with a row tile size of 128 (details in Appendix~\ref{appendix:eval-setup-details}).

\subsection{End-to-end Model Offline Prefill and Decode}
\label{subsec:eval:e2e}
\myparab{Single-GPU result.}
Fig.~\ref{fig:short-decode} shows the short-context decode-only throughput normalized against vLLM.
Compared to the highly specialized megakernel baselines, the \sysname framework achieves comparable performance, within $\hSingleShortDecodeOursVsVDCoresMin\times$–$\hSingleShortDecodeOursVsVDCoresMax\times$ of VDCores and $\hSingleShortDecodeOursVsTKAvg\times$ of ThunderKittens. 
For long-context decode, we outperform ThunderKittens by $\hSingleLongDecodeOursVsTKAvg\times$, while VDCores crashes due to KV cache index integer overflow (Fig.~\ref{fig:long-decode} in Appendix).
Note that ThunderKittens is a low-latency megakernel that is compatible only with Llama-3.2-1B at BS $=1$, and VDCores uses specialized schedules for each model at BS $\leq8$. The \sysname framework achieves a comparable throughput without manual tuning (\S\ref{subsec:graph-tune}) on the models and batch sizes they are hand-tuned for, demonstrating that our generality only comes at a minimal cost of performance. Appendix~\ref{sec:ext-eval-trace} analyzes where the \sysname framework gains and loses against VDCores and the non-overlapped version with a per-kernel trace breakdown analysis.

Compared to vLLM, a non-megakernel system, we achieve speedup consistently across all models and batch sizes, and on average obtain $\hSingleShortDecodeOursVsVllmAvg\times$ ($\hSingleShortDecodeOursVsVllmMin\times$–$\hSingleShortDecodeOursVsVllmMax\times$) higher short-context decode throughput than vLLM on H100. The trend holds on A100 and L40 (Appendix~\ref{sec:ext-eval-short-decode}), with average speedups of $\aSingleShortDecodeOursVsVllmAvg\times$ and $\lSingleShortDecodeOursVsVllmAvg\times$, respectively. 
As vLLM uses different kernels than us, to separate out this effect and measure the impact of overlap alone, we also run a version of \sysname with overlap turned off. The results show that overlap contributes significantly to the speedup: without it, the \sysname framework reaches $\hSingleShortDecodeNoovlpVsVllmAvg\times$, $\aSingleShortDecodeNoovlpVsVllmAvg\times$, and $\lSingleShortDecodeNoovlpVsVllmAvg\times$ of vLLM's throughput, and overlap adds an additional $\hSingleShortDecodeOursVsNoovlpAvg\times$, $\aSingleShortDecodeOursVsNoovlpAvg\times$, and $\lSingleShortDecodeOursVsNoovlpAvg\times$,  on H100, A100, and L40 GPUs, respectively. %
We achieve smaller speedups as model sizes grow because the kernel body execution time increases and dominates the overall latency. Appendix~\ref{sec:ext-eval} provides more detailed analysis.
In prefill settings, we only achieve comparable throughputs with baselines for a similar reason.

\begin{figure}[t]
    \centering
    \includegraphics[width=0.99\linewidth]{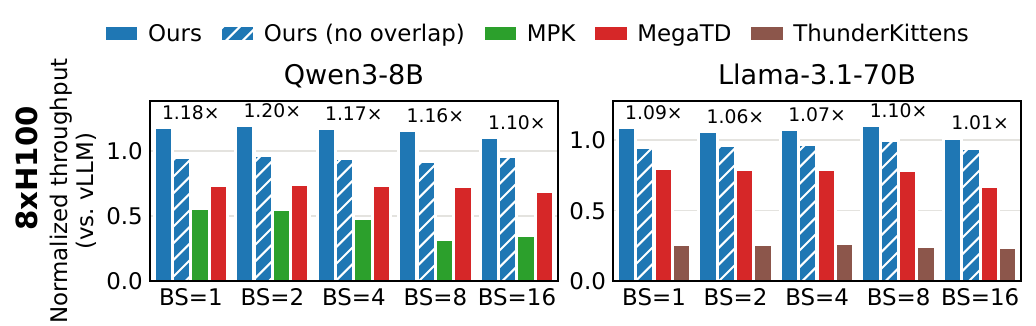}
    \caption{Long decode throughput. ``BS'' denotes global batch size. MPK does not support Llama-3.1-70B.}
    \label{fig:long-decode-multi}
\end{figure}
\myparab{Multi-GPU decode.} Fig.~\ref{fig:long-decode-multi} shows the long-context decode throughput on 8$\times$H100.
On average, our \sysname framework outperforms vLLM by $\hMultiLongDecodeOursVsVllmAvg\times$ ($\hMultiLongDecodeOursVsVllmMin\times$–$\hMultiLongDecodeOursVsVllmMax\times$).
The improvement is slightly lower than in the single-GPU decode, 
as computation in decode has only 1 wave with no room for the additional all-reduce operator to be hidden with overlap.
The megakernel compilers all achieve consistently lower throughput than vLLM in the batch sizes we evaluate, while \TK performs poorly as it is designed for larger batch sizes rather than our small-batch settings.

\begin{figure}[t]
    \centering
    \includegraphics[width=0.99\linewidth]{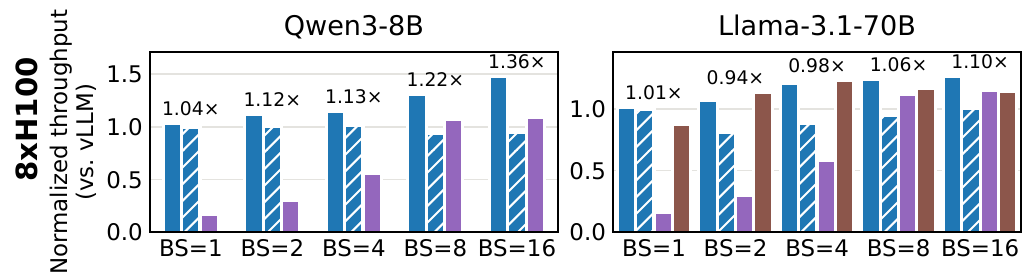}
    \caption{Prefill throughput. ``Ours'' uses AG-RS for BS $\geq$\hMultiLongPrefillAGRSFromBSQwenEight{} for Qwen3-8B and BS $\geq$\hMultiLongPrefillAGRSFromBSLlamaSeventy{} for Llama-3.1-70B, while AR is used for the rest of the BS. Legend is the same as Fig.~\ref{fig:long-decode-multi}.
}
    \label{fig:long-prefill-multi}
\end{figure}
\myparab{Multi-GPU prefill.} Unlike the single-GPU case, multi-GPU prefill benefits from overlap, as there are more computation kernel waves to hide communication. Fig.~\ref{fig:long-prefill-multi} shows that we outperform vLLM by $\hMultiLongPrefillOursVsVllmAvg\times$ ($\hMultiLongPrefillOursVsVllmMin\times$–$\hMultiLongPrefillOursVsVllmMax\times$) on 8$\times$H100. Compared to the non-overlapped version, we achieve $\hMultiLongPrefillOursVsNoovlpAvg\times$ higher throughput.
We find that the best parallel scheme depends on the batch size for each model. At BS=1, AG-RS reaches only $\hMultiLongPrefillOursAGRSVsOursARBSOneMin\times$–$\hMultiLongPrefillOursAGRSVsOursARBSOneMax\times$ of AR's throughput, while at BS=16 it is $\hMultiLongPrefillOursAGRSVsOursARBSSixteenMin\times$–$\hMultiLongPrefillOursAGRSVsOursARBSSixteenMax\times$ faster than AR. 
This demonstrates the importance of flexible programmability that allows for easy switching between parallel schemes depending on the workload.
Compared to the 8-GPU \TK megakernel on Llama-3.1-70B, our \sysname framework achieves $\hMultiLongPrefillOursVsTKMin\times$–$\hMultiLongPrefillOursVsTKMax\times$ of its throughput. Our major gain comes from a faster QKV projection and RoPE computation and better overlap between allgather and gate- and up-projection than \TK's broadcast-based allgather implementation, which demonstrates the importance of individual kernel tuning.

\subsection{Computation-communication Overlapping}
\label{subsec:eval:comm-comp-ovlp}
\begin{figure*}[t]
    \centering
    \includegraphics[width=\linewidth]{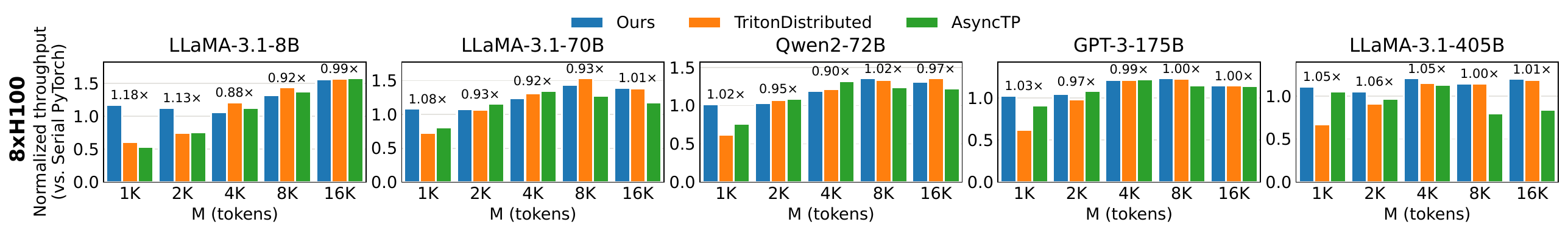}
    \caption{Allgather-GeMM overlap.}
    \label{fig:ag-gemm}
\end{figure*}
Fig.~\ref{fig:ag-gemm} shows that our green context design can overlap computation and communication efficiently. It achieves comparable performance to the best baselines in larger shapes where communication can be fully hidden. In smaller shapes, we achieve up to 18\% higher throughput because AsyncTP and TritonDistributed both use copy engine, which has a fixed exposed cost relatively high in small messages, while serial PyTorch uses NCCL for allgather, which does not fully utilize the GPU SMs and results in longer allgather time. 
Our framework automatically switches to kernel-based allgather for small shapes, and runs it on the computation green context partition to fully utilize GPU SMs.
This demonstrates the importance of switching kernels adaptively according to workloads. Our composability makes the switching seamless, which enables autotuning that can decide the best combination automatically.

\subsection{Usability Evaluation}
\label{subsec:eval:usability}

We quantify the user-side effort by counting the lines of code of our model implementations. 
We categorize each line of code into two main categories: base code and \sysname addition. The latter is further divided into six subcategories. Below are the classification rules:
\begin{packeditemize}
    \item \textbf{Base}: code irrelevant to \sysname, including kernel invocations with their arguments, buffer preparation such as tensor allocation, reshapes and initialization, backend dispatch, etc.
    \item \textbf{In}/\textbf{Out}: code preparing and passing the input/output semaphore arguments (\texttt{in\_semaphore}, \texttt{out\_semaphore}, map, target, and extents).
    \item \textbf{Alloc}: code constructing semaphore tensors, including choosing data tile sizes and sizing semaphore tensors.
    \item \textbf{Tune}: code invoking the autotuner and passing the resolved \texttt{tile\_config} to the launch.
    \item \textbf{Remap}: CTA remapping (\S\ref{subsec:reorder-prefetch}).
    \item \textbf{Other}: stream synchronization, symmetric-memory handling, profiling, and \texttt{tile\_config} bookkeeping.
\end{packeditemize}
Each line is assigned to one category, and a line mixing base and \sysname content is classified to \sysname categories conservatively.\footnote{Blank lines, comments, function signatures, assertions, and debug code are excluded. The large line count is due to our linter, which breaks down long lines into one argument per line.}
Table~\ref{tab:usability-loc} shows that our framework adds 2.1$\times$ lines per line of the original base code, with a total of 1767 additional lines on top of Base's 847 lines, confirming that only a small amount of code is needed to use the \sysname interface.
This code size can be further reduced by providing more helpers for common patterns, similar to calc\_target and calc\_extent (\S\ref{subsec:kernel-interface}).
For example, 0.29$\times$ lines of this are for autotuning, which contains repeated argument lists as the kernel calls; 0.20$\times$ are for allocation, which contains similar size calculation patterns.
The remaining is mostly for constructing the input and output mappings and can also be simplified with similar utility functions as those for target and extent. We leave these improvements to future work.

\subsection{Graph-level Tuning Ablation}
\label{subsec:eval:ablation}
\begin{figure}[t]
    \centering
    \includegraphics[width=0.9\linewidth]{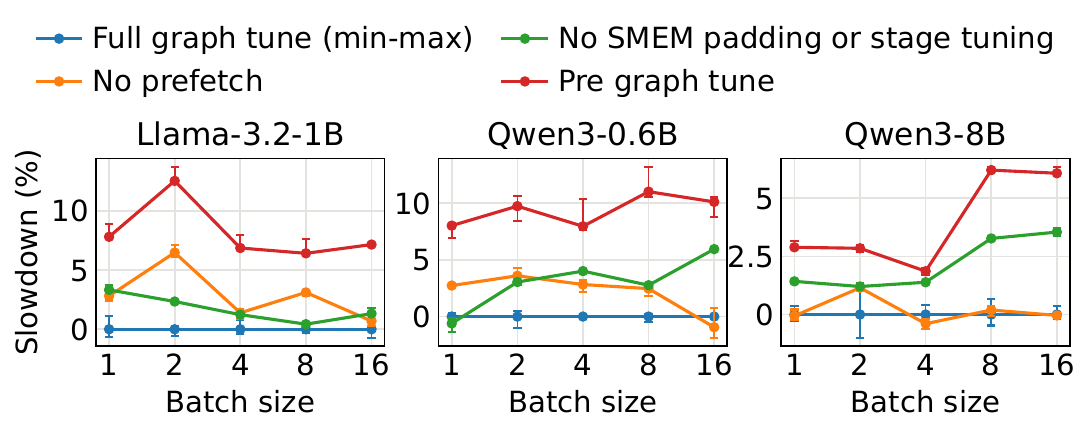}
    \caption{Effect of graph-level tuning. ``Pre graph tune'' does not run graph-level tuning. %
    The slowdown is relative to full graph tuning, with min and max plotted as error bars.}
    \label{fig:ablation-prefetch-smem}
\end{figure}
To evaluate the overlap-aware optimizations (\S\ref{sec:optimization}) and whether graph-level tuning is necessary to make them effective, we rerun the short-context decode on H100 with graph tuning entirely skipped, and with one optimization removed from the graph tuning search at a time while still tuning the rest. Fig.~\ref{fig:ablation-prefetch-smem} shows the slowdown of the per-step decode time relative to the full graph tuning.
Without graph tuning, the \sysname framework uses kernel configurations autotuned on each kernel in isolation, and the decode step becomes more than 10\% slower in certain cases, confirming the effectiveness of graph-level tuning.

We next ablate the effect of the proposed knobs in the graph-level tuning. Without tuning prefetching, the decode step slows down by 2-6\% for small models, where GeMMs are small and the prefetching covers a larger fraction of the total weight tiles.
Disabling both SMEM padding and stage tuning slows down the decode step by up to 6\%. %
Appendix~\ref{sec:ext-eval-remap} further shows that disabling CTA remapping slows down the multi-GPU prefill by up to 1.32$\times$.

\begin{table}
  \centering
  \caption{Model forward pass lines of code breakdown. The ratio row divides each column by the Base lines.
  }
  \label{tab:usability-loc}
  \footnotesize
  \setlength{\tabcolsep}{3pt}
  \begin{adjustbox}{max width=0.83\columnwidth}
  \begin{tabular}{lrrrrrrrr}
    \toprule
     & \multirow{2}{*}{Base} & \multicolumn{7}{c}{\sysname addition} \\ \cmidrule(lr){3-9}
     &  & In & Out & Alloc & Tune & Remap & Other & Total \\
    \midrule
    Lines & 847 & 490 & 611 & 171 & 244 & 111 & 140 & 1767 \\
    Ratio & 1.00 & 0.58 & 0.72 & 0.20 & 0.29 & 0.13 & 0.17 & 2.09 \\
    \bottomrule
  \end{tabular}
  \end{adjustbox}
\end{table}

\subsection{Composability Case Study}
\label{subsec:eval:casestudy}

As mentioned in \S\ref{subsec:kernel-interface}, as long as two kernels are adapted to the same data tiling scheme, they can compose freely. 
To demonstrate this concretely, we illustrate with how two different MLP TP strategies, AG-RS and allreduce (AR), can share the same code path of the MLP computation (pseudocode in Listing~\ref{lst:agrs-ar} in Appendix). The AG-RS path runs AG$\rightarrow$MLPCompute$\rightarrow$RS, while the allreduce (AR) path runs NoOP$\rightarrow$MLPCompute$\rightarrow$AR.
By all using the row-wise data tiling scheme for the intermediate tensors, namely the input and output of AG, MLPCompute, RS, and AR, the same MLPCompute function can be shared between the two communication paths. %
This allows us to easily switch between AG-RS and AR without changing other parts of the code path, demonstrating the composability of our system.

In fact, we add AR after the AG-RS path is implemented, and adding AR for both MLP and Attention blocks only required us to edit in total 346 Python LoC in the user program, including weight sharding logic, control-flow branching logic, and safety checks. 

%% file: sections/related-work.tex
\section{Related Work}
Several lines of work study how to reduce the overhead in the traditional \kernelperop scheme.
Megakernels pack individual kernels together to reduce overhead. Some works develop highly specialized megakernels for certain workloads or hardware, such as \TK megakernels, FlashMoE, Perseus, and UniEP~\cite{megatklat,megatktput,flashmoe,perseus,uniep,megamoetk}. 
To improve portability, several works design general frameworks to generate megakernels. MPK, Event Tensor, MegaTritonKernels, and TileRT can compile a user program resembling PyTorch to a megakernel~\cite{cheng2026mpk,etc,zheng2026ditron,tilelink,tilert}. VDCores~\cite{vdcores} proposes a fine-grained megakernel programming model where users schedule the compute or memory operations per SM. However, they all fuse into one monolithic kernel and lose interoperability and debuggability. Our \sysname framework retains these properties while achieving comparable performance to megakernels.

In multi-GPU settings, computation-communication overlap is another approach to reducing the \kernelperop overhead.
Most of them design specific fused kernels or overlap schemes for certain operator combinations, such as AG+GeMM, GeMM+RS, GeMM+AR, Dispatch+GroupGeMM, and GroupGeMM+Combine~\cite{flashoverlap,googletp,asynctp,chang2024flux,zhang2025comet,deepep2025,fabriclib,ringattention,jacobs2023deepspeed}. CoCoNet, TritonDistributed, Mercury, Syncopate, and ParallelKittens design domain-specific languages (DSLs) for generating customized kernel overlap or fusion~\cite{tritondist,qiang2026syncopate,mercury,sul2026parallelkittens,coconet}.
They all focus on co-designing computation and communication to achieve high performance but often introduce kernel coupling. We instead focus on a general interface for overlapping any number of both computation and communication kernels with minimal coupling.

Our interface is closest to CuSync and \ETC, which also let users express tile-level dependencies~\cite{cusync,etc}. CuSync builds dependencies on CTA IDs, so a consumer kernel must know which producer CTAs produce the tile it loads. This couples the tilings of the two kernels. \ETC represents a dependency as an \emph{event} that counts the completion of a set of producer tasks. Users map the task coordinates of both kernels onto a shared event tensor. Since an event records that tasks have finished rather than that a region of a tensor is ready, the event layout is a convention that each pair of kernels must agree on. \sysname instead ties every semaphore to a data tile of the tensor itself, so the data tiling alone defines the contract between any two kernels. \ETC also fuses its device functions into a single persistent kernel, so closed-source kernels cannot participate and per-kernel profiling tools no longer apply, whereas \sysname launches each operator as a separate kernel and overlaps them at runtime.

%% file: sections/conclusion.tex
\section{Conclusion}
ML frameworks face a choice between the modularity of the \kernelperop paradigm and the performance of megakernels. We observed that megakernels gain their performance by removing kernel barriers, which does not require fusing kernels. \sysname runs each operator as a separate kernel, launches dependent kernels on separate CUDA streams, and synchronizes them on tiles of the tensors that connect them. Since kernels synchronize on data instead of on each other's CTAs, users can compose and reuse kernels freely and debug each operator in isolation. \sysname matches the throughput of hand-tuned ThunderKittens megakernels on their target workloads, showing that ML frameworks can reach megakernel performance without fusing operators and without giving up the modularity of the \kernelperop paradigm.

\section*{Acknowledgement}
We thank Riley Bakes for implementing a multi-threaded version of semaphore signal and wait functions and Chi-Sheng Liu for an early exploration of the SM scheduling issue.

%% file: sections/appendix.tex
\newpage
\appendix
\section{Automatic Memory Management Details}
\label{app:mem}

\paragraph{Tracking peer progress.} Before reusing a symmetric memory tensor, all peers must have passed the $N$-kernel mark to ensure it is safe to reuse globally. To track peer progress deterministically and without involving the host, we ask the user to mark kernels that produce a symmetric memory tensor, and kernels that consume data of such a tensor where the consumed data requires the contribution of all peers. These markers expose the dependency between local and peer kernels, allowing us to track peer progress, because the completion of such a consumer kernel implies the successful launch (and at least partial execution) of the peer producer kernels. Combined with the $N$-kernel limit, we know the peers have progressed at least up to the $(N-1)$-th kernel before the producer kernel. We thus add a second tensor reference to estimate the peer's progress using the above method.

\paragraph{Deterministic allocation.}
Finally, to guarantee determinism, we develop a preliminary approach where we use a dedicated memory pool for only symmetric tensors. The pool only reuses when the allocation size matches exactly the tensor in the pool, and returns the earliest free tensor for simplicity and determinism. 
We assume a SPMD-style programming where all peers run the same code to remove sources of inconsistencies in symmetric memory allocation. 
However, Python's garbage collection (GC) may still introduce inconsistent memory release timings. We address it by asking users to manually run GC periodically (\eg once every layer) so that all peers run at synchronized timings. To reduce frequent GC overhead, we recommend that the user freeze the currently tracked objects via \texttt{gc.freeze()} before running the model forward so that later GC only scans new objects created during model forward. This freeze technique is applied in our dense model forward evaluation (\S\ref{sec:eval-setup}).

\section{Implementation Details}
\label{app:impl}
\paragraph{Supported kernels.}
We integrate most of our computation kernels from existing open-source projects, including GeMM, normalization, elementwise, and attention kernels from TritonDistributed, and the unified attention kernel from vLLM. We added prefetch and Stream-K optimizations on top of TritonDistributed's GeMM kernel. Communication kernels are implemented on our own, including allgather, reducescatter, distributed transpose~\cite{megatktput} and allreduce kernels, which rely on peer memory access or multicast memory.
Closed-source kernels include GeMM kernels from cuBLASLt, attention kernels from cuDNN and FlashAttention, and copy-engine-based kernels for allgather and reducescatter.\footnote{FlashAttention is open-source but in CUDA, and we only support Triton currently. We thus treat it as closed-source for integration.}

\paragraph{Optimization for closed-source kernel scheduling.}
Synchronization kernels around subkernels use a complementary backend against the subkernel's to avoid interference and achieve more timely synchronization. 
When the subkernel uses copy engines, we use ordinary kernels; if subkernels are ordinary kernels, we use stream memory operations~\cite{streamops}, which is feasible as the synchronization is simply synchronizing at one variable, which is supported by stream operations.
We empirically observe that using the same backend for both will contend for the same resources and get serialized. For example, as we use two streams for scheduling subkernels, the second subkernel on the second stream may run before the synchronization signal kernel of the first subkernel, and serializing and delaying the signal until the end of the second subkernel.
Using a different backend addresses the issue, and we hypothesize that it is because stream operations use complementary hardware resources to ordinary kernels.

\begin{lstlisting}[mathescape=true,caption={Illustrative example showing composability enables allgather-reducescatter and allreduce to share the code path. Semaphore mappings and targets are omitted.},float=t,tabsize=2,label=lst:agrs-ar]
def TP_MLP(x, S_x):
 # S_x:          *row-wise plain tiling*
 if AG_RS:
   S_pre = ... #  *row-wise plain tiling*
   x_pre = allgather(x, S_x, S_pre, ...)
 else: # no-op, pass through
   S_pre = S_x
   x_pre = x
 x_d, S_d = MLPCompute(x_pre, S_pre)
 if AG_RS:
   S_mlp = ... # *row-wise plain tiling*
   x_mlp = reducescatter(x_d,S_d,S_mlp,...)
 else:
   S_mlp = ... # *row-wise plain tiling*
   x_mlp = allreduce(x_d, S_d, S_mlp, ...)
 return x_mlp, S_mlp

def MLPCompute(x, S_x): # shared code path
 # S_x:          *row-wise plain tiling*
 # gate and up projection
 S_u = ...
 x_u = matmul(x, S_x, S_u, ...)
 # silu_mul
 ...
 # down projection
 S_d = ... #     *row-wise plain tiling*
 x_d = matmul(x_silu, S_silu, S_d, ...)
 return x_d, S_d
\end{lstlisting}

\section{Extended Evaluation}
\label{sec:ext-eval}
\subsection{Extended Evaluation Setup}
\label{appendix:eval-setup-details}
\paragraph{Baselines details.} We compare against the serving framework \emph{vLLM} (v0.31.0), \emph{TritonDistributed} (commit hash 12c2389), a distributed compiler designed for computation-communication overlapping, megakernel compilers \emph{MPK} (commit hash 8b981a4)
and MegaTritonKernel~\cite{megatd} (referred to as \emph{MegaTD}) from TritonDistributed, 
megakernel programming framework \emph{VDCores} (commit hash fd75b8d), and specialized handwritten megakernels \emph{ThunderKittens} including both the single-GPU low-latency and the 8-GPU high-throughput versions.
For clarity, we omit MegaTD, VDCores, and the single-GPU ThunderKittens in prefill plots as they only support decode. MPK is also omitted in prefill as it is over an order of magnitude slower than ours or vLLM. In decode, we omit TritonDistributed as its AG-RS parallel scheme incurs high overhead and makes it much slower than others.
For accurate comparison, we extract the forward function vLLM uses and record its execution as a CUDA graph and wire it into the same token greedy sampler we implement to eliminate overheads caused by host and framework processing differences, which are observed to be up to 17\% and 400\% for decode and prefill on small models, respectively.
For communication-computation overlap we compare against \emph{AsyncTP}~\cite{asynctp} and TritonDistributed.

\paragraph{Data tiling implementation details}
Most intermediate tensors use a row-wise data tiling scheme, where each data tile covers 128 consecutive rows across the full hidden dimension, regardless of the producer kernel's tile size.
There are a few exceptions in the AR scheme, where we use two-dimensional data tiles so that consumers can start on partially produced columns, including the output of the gate and up projection, silu activation, and the self-attention output.

\subsection{Short-context Decode on A100 and L40}
\label{sec:ext-eval-short-decode}
\begin{figure*}
    \centering
    \includegraphics[width=0.99\linewidth]{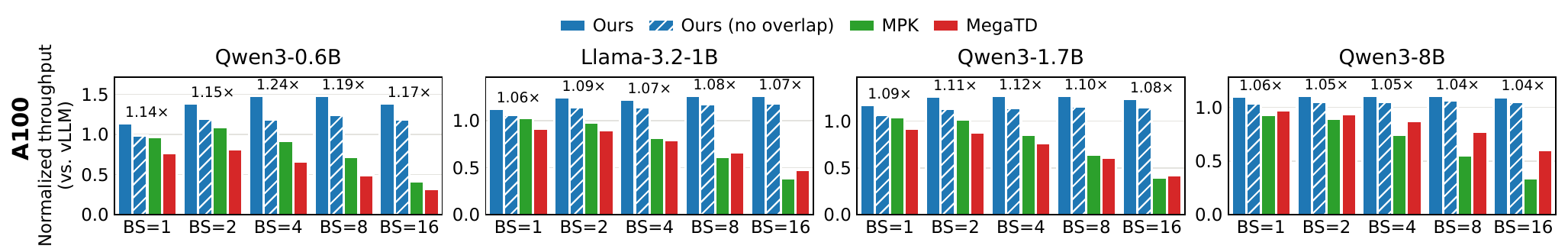}
    \includegraphics[width=0.99\linewidth]{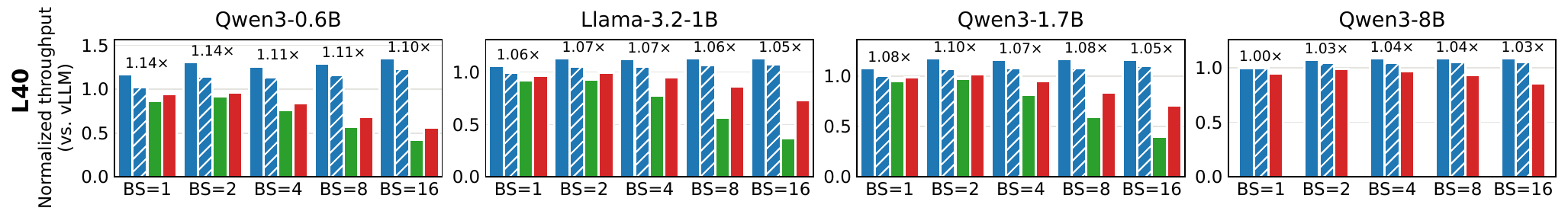}
    \caption{Short decode throughput on A100 (top) and L40 (bottom) using 128 prompt tokens and 64 output tokens. Numbers above bars indicate the speedup of ``Ours'' over the best of the other systems. VDCores and \TK do not support GPUs prior to Hopper.}
    \label{fig:short-decode-a100-l40}
\end{figure*}
Fig.~\ref{fig:short-decode-a100-l40} shows the short-context decode throughput on A100 and L40. The \sysname framework achieves $\aSingleShortDecodeOursVsVllmAvg\times$ ($\aSingleShortDecodeOursVsVllmMin\times$–$\aSingleShortDecodeOursVsVllmMax\times$) and $\lSingleShortDecodeOursVsVllmAvg\times$ ($\lSingleShortDecodeOursVsVllmMin\times$–$\lSingleShortDecodeOursVsVllmMax\times$) higher throughput than vLLM on A100 and L40, respectively.

\subsection{Short-Context Decode MBU Analysis}
\begin{figure*}
    \centering
    \includegraphics[width=0.99\linewidth]{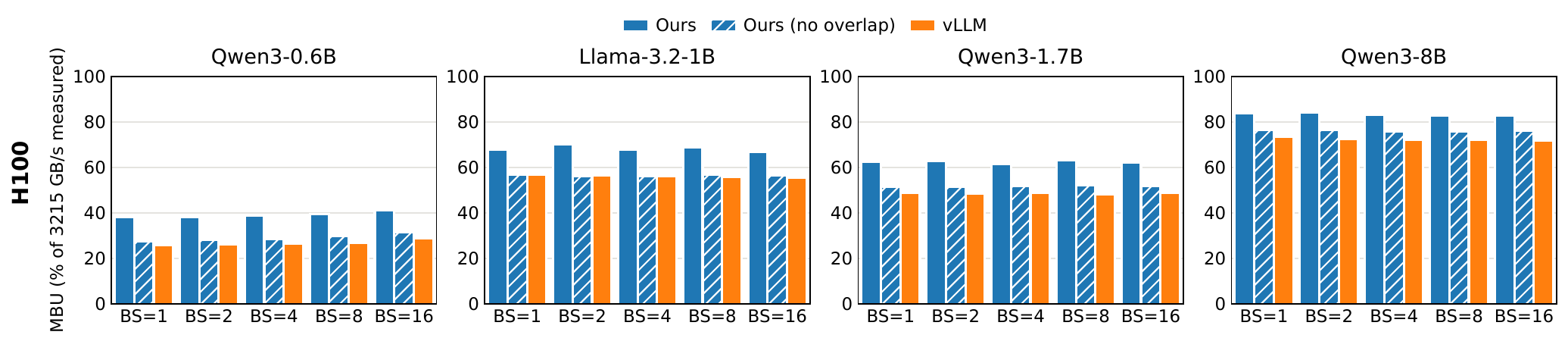}
    \includegraphics[width=0.99\linewidth]{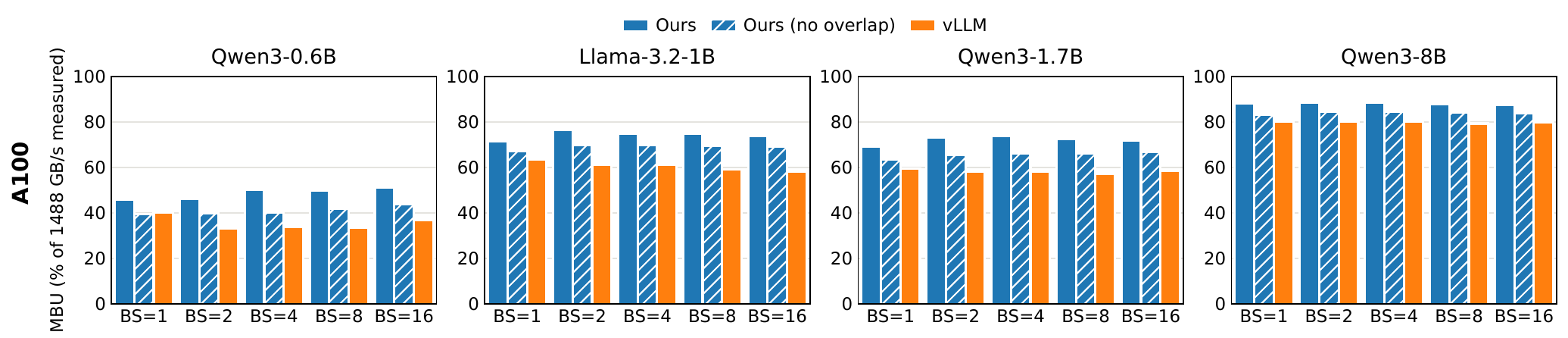}
    \includegraphics[width=0.99\linewidth]{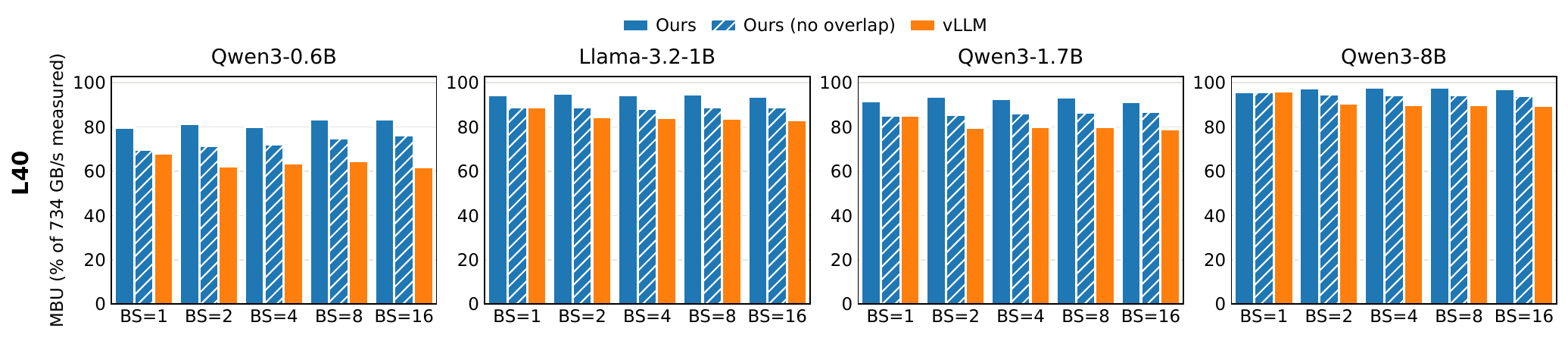}
    \caption{Short-context decode MBU analysis. The MBU is calculated against the maximum achievable bandwidth measured using \texttt{nvbandwidth} on the same GPU, instead of the theoretical maximum from hardware specifications.}
    \label{fig:decode-mbu}
\end{figure*}
We use the model weights and KV cache size to estimate the total memory read traffic volume, and derive the bandwidth by dividing with execution time. We then measure the maximum achievable bandwidth using \texttt{nvbandwidth} on the same GPU, and calculate the memory bandwidth utilization (MBU) as the ratio of the two.
Fig.~\ref{fig:decode-mbu} shows the MBU of short decode on A100 and L40. vLLM achieves a high MBU on larger models, leaving little room for \sysname to improve. L40 has a lower achievable memory bandwidth, and thus the \sysname framework has lower speedups on L40 than A100 in general.

\subsection{Long-context Decode and Prefill}
\begin{figure*}
    \centering
    \includegraphics[width=0.99\linewidth]{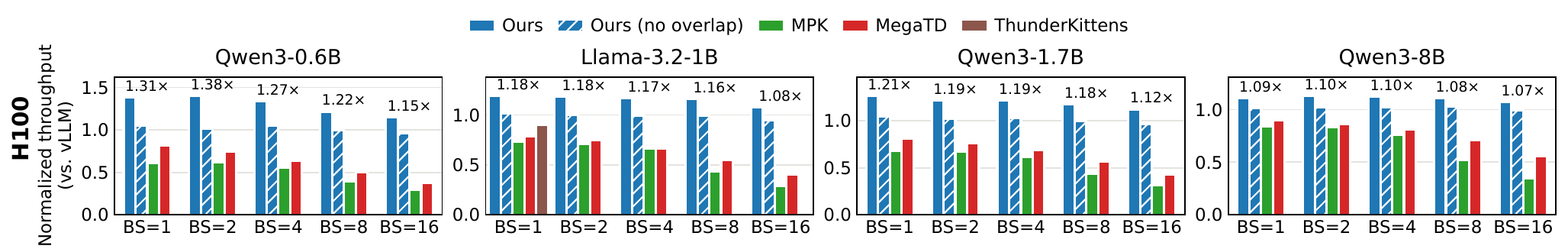}
    \includegraphics[width=0.99\linewidth]{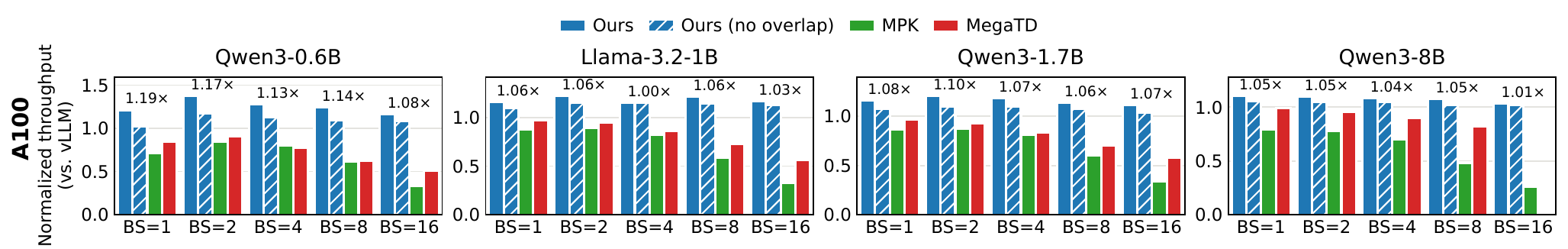}
    \caption{Long decode throughput using 1024 prompt tokens and 1024 output tokens. VDcores crashes due to KV cache index integer overflow.}
    \label{fig:long-decode}
\end{figure*}

Fig.~\ref{fig:long-decode} shows single-GPU decode throughput with a 1024-token prompt and output length on H100 and A100. ThunderKittens shows lower performance than vLLM while the \sysname framework consistently outperforms vLLM on both H100 and A100, achieving on average 
$\hSingleLongDecodeOursVsVllmAvg\times$ 
($\hSingleLongDecodeOursVsVllmMin\times$–$\hSingleLongDecodeOursVsVllmMax\times$) and  $\aSingleLongDecodeOursVsVllmAvg\times$ ($\aSingleLongDecodeOursVsVllmMin\times$–$\aSingleLongDecodeOursVsVllmMax\times$) higher throughput than vLLM on H100 and A100, respectively.

\begin{figure*}
    \centering
    \includegraphics[width=0.99\linewidth]{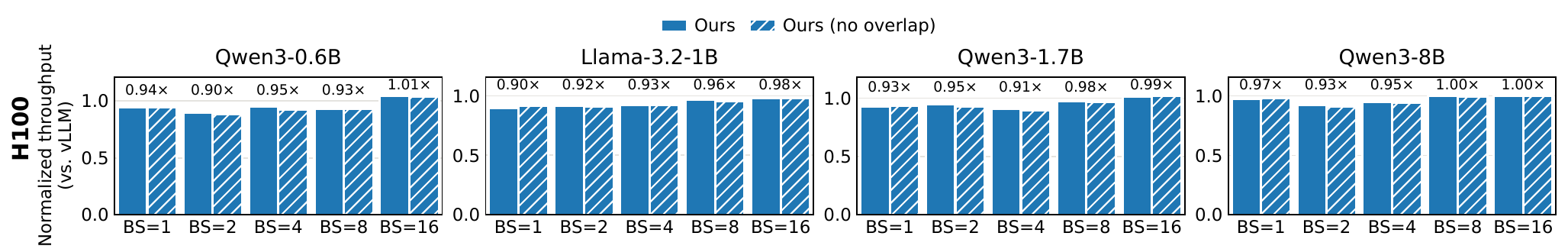}
    \includegraphics[width=0.99\linewidth]{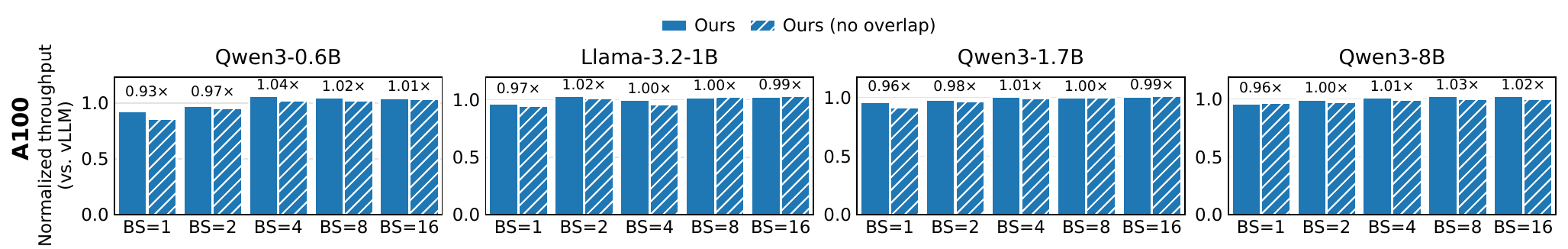}
    \caption{Prefill throughput using 1024 prompt tokens.}
    \label{fig:prefill}
\end{figure*}

\subsection{Multi-GPU Results on 4$\times$A100}
\label{sec:ext-eval-multi-a100}
\begin{figure}[t]
    \centering
    \includegraphics[width=\linewidth]{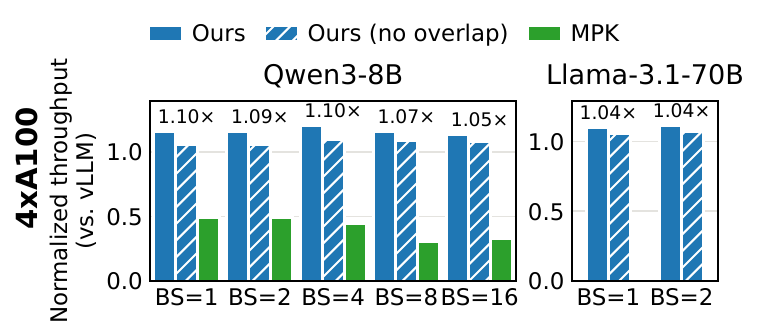}
    \caption{Long decode throughput on 4$\times$A100. ``BS'' denotes global batch size. MPK does not support Llama-3.1-70B, and BS$>$2 runs out-of-memory for all systems on Llama-3.1-70B.}
    \label{fig:long-decode-multi-a100}
\end{figure}
\begin{figure}[t]
    \centering
    \includegraphics[width=\linewidth]{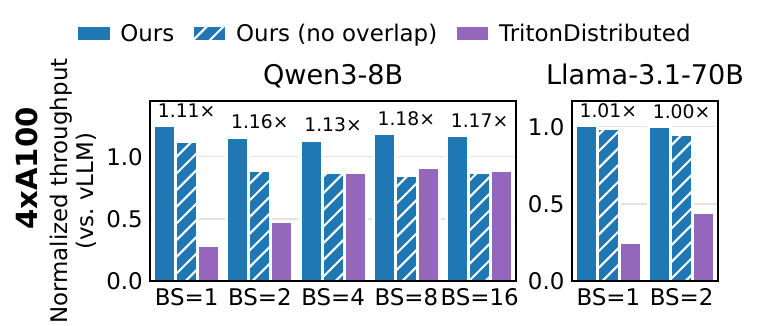}
    \caption{Prefill throughput on 4$\times$A100. BS$>$2 runs out-of-memory for all systems on Llama-3.1-70B.}
    \label{fig:long-prefill-multi-a100}
\end{figure}
Fig.~\ref{fig:long-decode-multi-a100} and Fig.~\ref{fig:long-prefill-multi-a100} show the long-context decode and prefill throughput on 4$\times$A100 with tensor parallelism, which follow similar trends as on 8$\times$H100 (\S\ref{subsec:eval:e2e}).
For decode, the \sysname framework outperforms vLLM by $\aMultiLongDecodeOursVsVllmAvg\times$ ($\aMultiLongDecodeOursVsVllmMin\times$–$\aMultiLongDecodeOursVsVllmMax\times$) on average.
For prefill, the \sysname framework outperforms vLLM by $\aMultiLongPrefillOursVsVllmAvg\times$ ($\aMultiLongPrefillOursVsVllmMin\times$–$\aMultiLongPrefillOursVsVllmMax\times$) and the non-overlapped version by $\aMultiLongPrefillOursVsNoovlpAvg\times$.
The best parallelization scheme also depends on the batch size: on Qwen3-8B, at BS $=1$, AG-RS reaches only $\aMultiLongPrefillOursAGRSVsOursARBSOneMin\times$ of AR's throughput, while at BS $=16$ it is $\aMultiLongPrefillOursAGRSVsOursARBSSixteenMin\times$ faster than AR.

\begin{figure*}
    \centering
    \includegraphics[width=\linewidth]{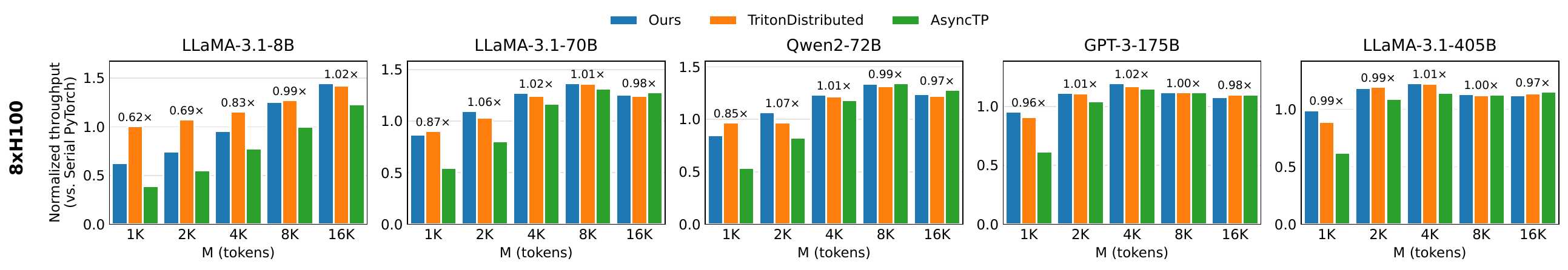}
    \caption{GeMM-reducescatter overlap on 8$\times$H100. Numbers above bars indicate the speedup of ``Ours'' over the best of the other systems.}
    \label{fig:gemm-rs}
\end{figure*}

\subsection{CTA Remapping Ablation}
\label{sec:ext-eval-remap}
\begin{table}[t]
  \footnotesize
  \centering
  \caption{Prefill latency (ms) with and without CTA remapping on 8$\times$H100, using the AG-RS parallelization scheme and 1024-token prompts.}
  \label{tab:noremap-prefill}
  \begin{tabular}{c|ccc}
    \toprule
    Model & BS & Remap & No remap (slowdown) \\
    \midrule
    \multirow{2}{*}{\makecell{Llama-3.1\\-70B}} & 8 & 258.7 ms & 303.3 ms (1.17$\times$) \\
     & 16 & 503.2 ms & 572.3 ms (1.14$\times$) \\
    \midrule
    \multirow{2}{*}{\makecell{Qwen3\\-8B}} & 8 & 41.4 ms & 49.4 ms (1.19$\times$) \\
     & 16 & 67.2 ms & 88.9 ms (1.32$\times$) \\
    \bottomrule
  \end{tabular}
\end{table}
Table~\ref{tab:noremap-prefill} shows the multi-GPU prefill latency with and without CTA remapping (\S\ref{subsec:reorder-prefetch}) in the AG-RS scheme. Without remapping, GeMM CTAs are launched in their default order that does not match the all-gather's order, causing delays.
The result shows that the prefill latency is increased by 1.14$\times$--1.17$\times$ on Llama-3.1-70B and 1.19$\times$--1.32$\times$ on Qwen3-8B without remapping, confirming the benefit of remapping.

\subsection{GeMM-Reducescatter Overlap }
Fig.~\ref{fig:gemm-rs} shows the GeMM-reducescatter overlap benchmark from TritonDistributed, which runs the MLP down projection followed by reducescatter under tensor parallelism, using the same models for the weight shapes as the allgather-GeMM benchmark (\S\ref{subsec:eval:comm-comp-ovlp}).
For M $\geq$ 2K tokens, the \sysname framework performs on par with the best baselines on all models except Llama-3.1-8B, at 0.97$\times$--1.07$\times$ of their throughput.
At smaller shapes, the \sysname framework falls behind TritonDistributed, reaching 0.62$\times$--0.83$\times$ of its throughput on Llama-3.1-8B at M $\leq$ 4K and 0.85$\times$--0.87$\times$ on Llama-3.1-70B and Qwen2-72B at M $=$ 1K. This is because TritonDistributed switches to a different GeMM kernel that directly stores result to peer memory, while the \sysname framework always uses the copy engine regardless of shapes, which has higher overhead in smaller shapes. We could implement a kernel-based reducescatter to replace copy engine similar to our allgather implementation. We leave it for future work.

\subsection{Trace Analysis of Overlap Benefits}
\label{sec:ext-eval-trace}
\begin{figure}[t]
    \centering
    \includegraphics[width=\linewidth]{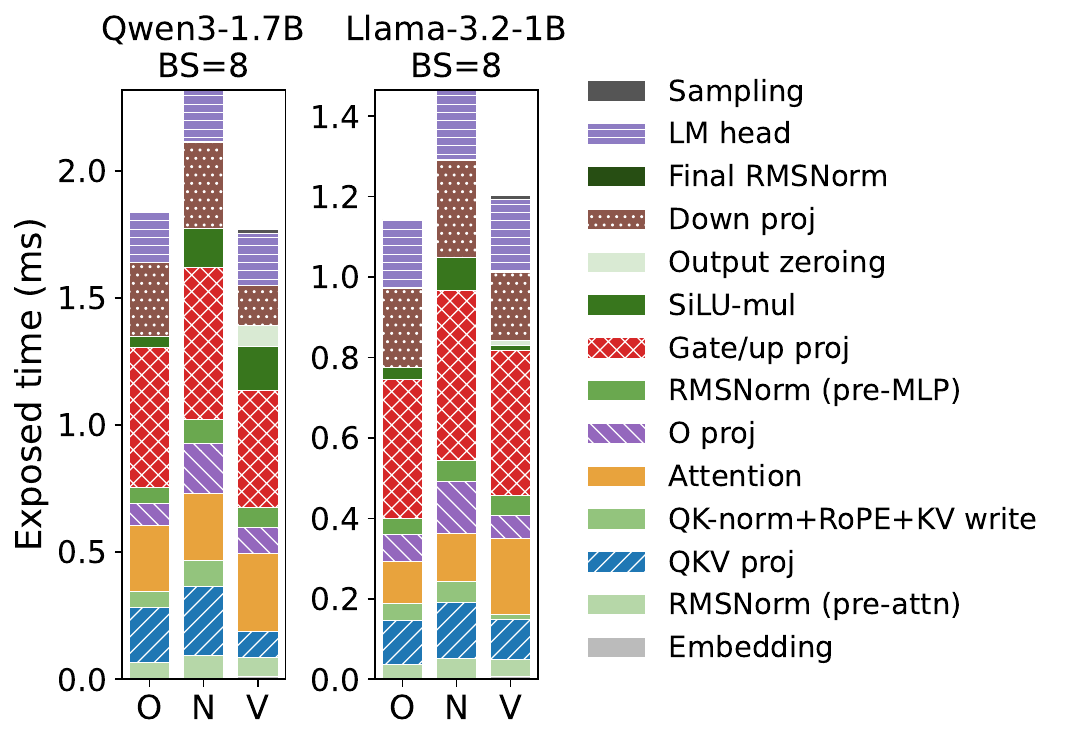}
    \caption{Exposed time per operator of one short-context decode step at BS $=8$ on H100. O, N, V denote \sysname, \sysname without overlap, and VDCores, respectively.}
    \label{fig:vdcores-breakdown}
\end{figure}
To understand where the \sysname framework gains and loses against non-overlapped executions and megakernels, we profile one short-context decode step at BS $=8$ on H100 with intra-kernel tracing for \sysname, \sysname without overlap, and VDCores, and attribute time to the earliest finishing operator to account for overlap (Fig.~\ref{fig:vdcores-breakdown}). In other words, a later kernel does not count the overlapped portion with earlier kernels, but only the exposed portion after all prior kernels finish.
Compared to the no-overlap version, overlap removes most of the exposed time of the small operators between GeMMs, such as SiLU and RMSNorms, and prefetching reduces the GeMM time as the loading is overlapped with prior kernels, most notably for the output projection.
Compared to VDCores, the \sysname framework exposes on par or less time on the small operators, while the GeMMs are mixed.
On the other hand, our attention is faster on both models, which on Llama-3.2-1B makes \sysname faster overall. %
This result shows that the \sysname framework can achieve similar performance optimizations as a megakernel. It also shows aside from overlapping, achieving high performance for individual kernels is also important. Our system can reuse existing kernels and easily switch between them, allowing kernels to be independently evolved and optimized. This makes achieving high performance for individual kernel much easier than in a megakernel, where the kernel is often coupled with the rest of the system and reusing other kernels is difficult.

\subsection{Trace Breakdown}
\begin{figure*}
    \centering
    \includegraphics[width=0.99\linewidth]{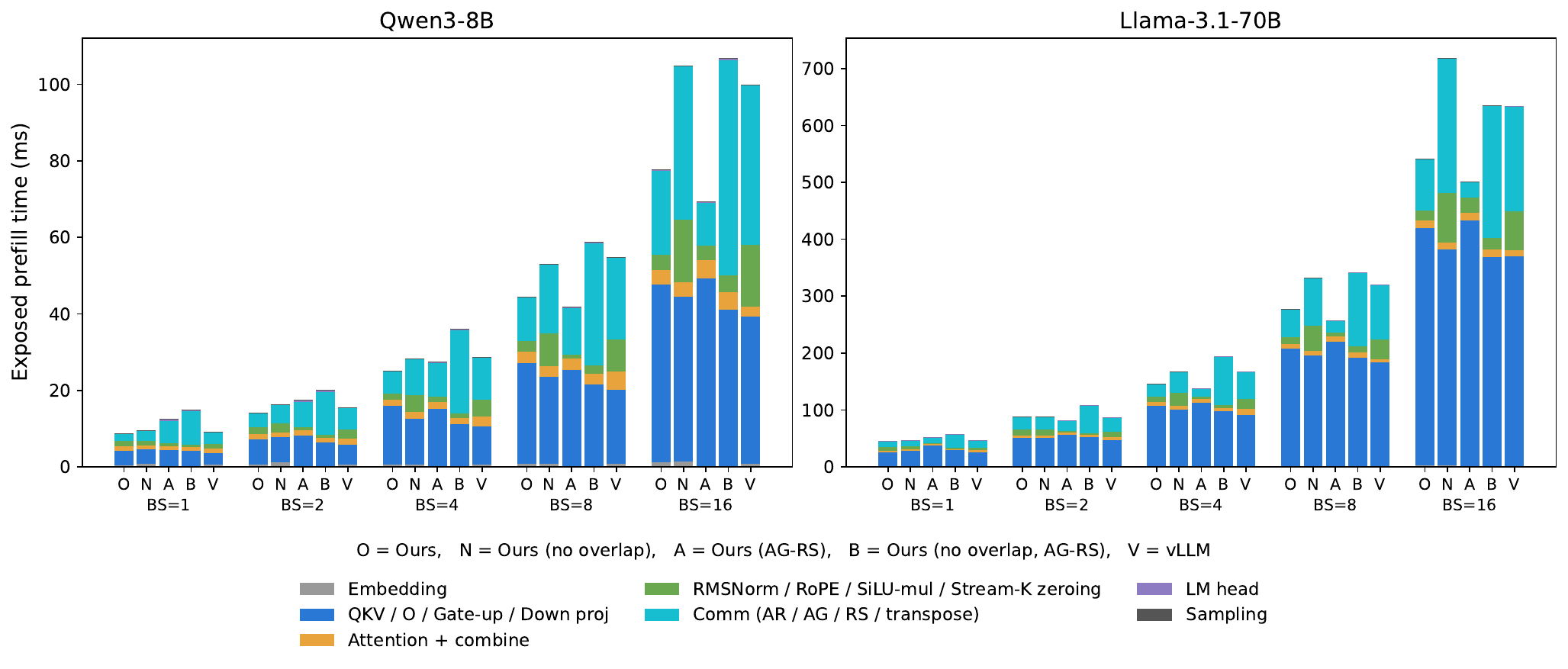}
    \caption{Exposed time per operator group of the 1024-token prefill step on 8$\times$H100. O, N, A, B, V denote \sysname, \sysname without overlap, \sysname with AG-RS, \sysname with AG-RS and without overlap, and vLLM, respectively.}
    \label{fig:trace-breakdown-prefill}
\end{figure*}

\begin{figure*}
    \centering
    \includegraphics[width=0.99\linewidth]{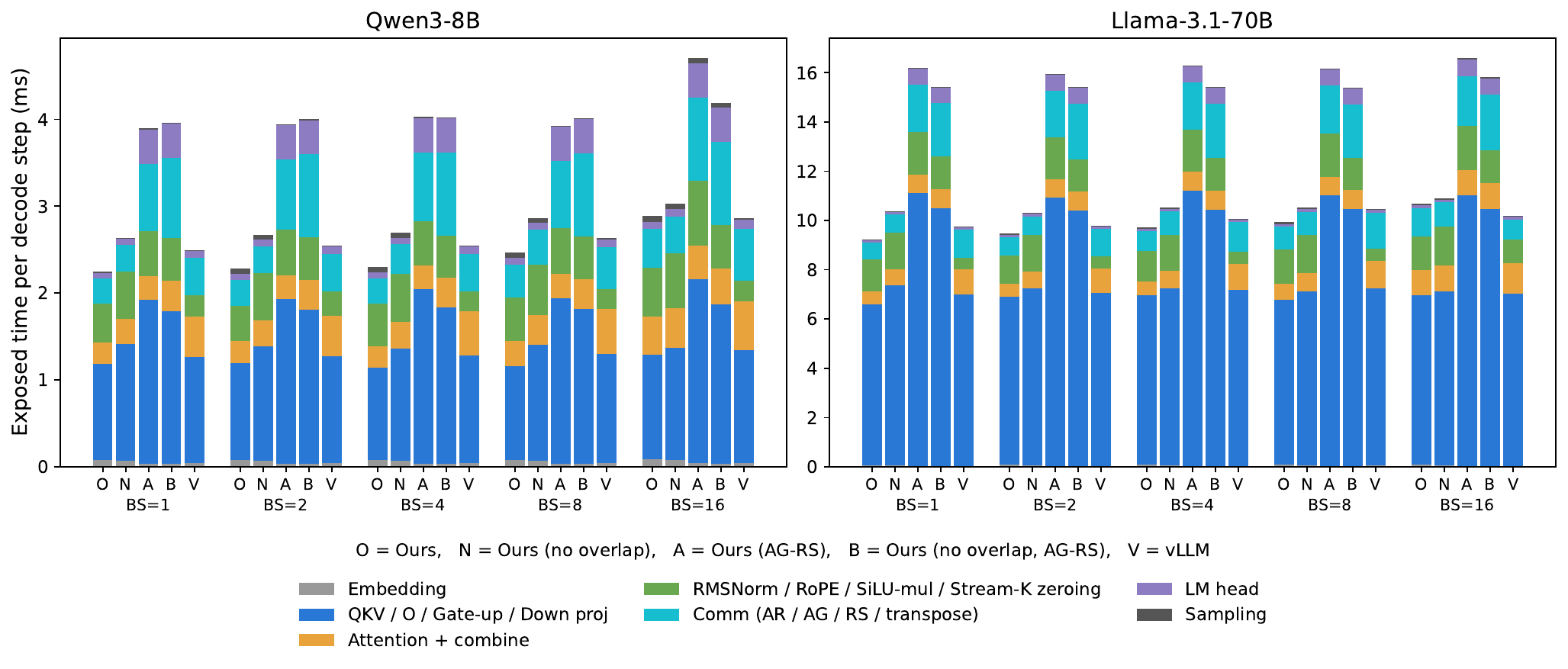}
    \caption{Exposed time per operator group of one long-context decode step on 8$\times$H100. O, N, A, B, V denote \sysname, \sysname without overlap, \sysname with AG-RS, \sysname with AG-RS and without overlap, and vLLM, respectively. vLLM fuses the allreduce with the following RMSNorm at BS $\leq 8$, which is counted as communication.}
    \label{fig:trace-breakdown-decode}
\end{figure*}

To understand where the multi-GPU gains come from, we profile the long-context prefill and decode on 8$\times$H100 and attribute the time to the operator that is exposed, as in Appendix~\ref{sec:ext-eval-trace} (Fig.~\ref{fig:vdcores-breakdown}).

\paragraph{Prefill.} Fig.~\ref{fig:trace-breakdown-prefill} shows the prefill breakdown. We focus on analyzing AG-RS. As shown, the gain primarily comes from better overlap between computation and communication.
Compared to AG-RS without overlap and vLLM, which do not overlap communication with GeMMs, AG-RS with overlap hides most of its communication (observe the large reduction in cyan bars from ``B'' to ``A''), with only a small amount of impact on the GeMM performance. At large BS, this overlap pays off and hides more communication than the AR scheme, making AG-RS the fastest system.

\paragraph{Decode.} Fig.~\ref{fig:trace-breakdown-decode} shows the decode breakdown. Unlike prefill, the gain in decode mainly comes from reduced GeMM and small operator times, consistent with the observation in Fig.~\ref{fig:vdcores-breakdown}. In this small shape GeMMs finish in one wave, leaving no room to hide the AR communication.  

\clearpage